\documentclass[iop,revtex4]{emulateapj}
\usepackage{amsmath}
\usepackage{rotating}
\usepackage{lscape}

\newcommand{\myemail}{paola.santini@oa-roma.inaf.it}

\slugcomment{To appear in {\it The Astrophysical Journal}}

\shorttitle{Stellar masses from the CANDELS survey}
\shortauthors{P. Santini et al.}

\begin{document}


\title{Stellar masses from the CANDELS survey: the GOODS-South and UDS  fields}

\author{
P. Santini\altaffilmark{1},
H.~C.~Ferguson\altaffilmark{2},
A.~Fontana\altaffilmark{1},
B.~Mobasher\altaffilmark{3},
G.~Barro\altaffilmark{4},
M.~Castellano\altaffilmark{1},
 S.~L.~Finkelstein\altaffilmark{5},
A.~Grazian\altaffilmark{1},
L.~T.~Hsu\altaffilmark{6},
B.~Lee\altaffilmark{7},
S.-K.~Lee\altaffilmark{8},
J.~Pforr\altaffilmark{9},
M.~Salvato\altaffilmark{6},
T.~Wiklind\altaffilmark{10},
S.~Wuyts\altaffilmark{6},
O.~Almaini\altaffilmark{11},
M.~C.~Cooper\altaffilmark{12}, 
A.~Galametz\altaffilmark{6},
B.~Weiner\altaffilmark{13},
R.~Amorin\altaffilmark{1},
K.~Boutsia\altaffilmark{1},
C.~J.~Conselice\altaffilmark{14},
T.~Dahlen\altaffilmark{2},
M.~E.~Dickinson\altaffilmark{9},
M.~Giavalisco\altaffilmark{7},
N.~A.~Grogin\altaffilmark{2},
Y.~Guo\altaffilmark{4},
N.~P.~Hathi\altaffilmark{15},
D.~Kocevski\altaffilmark{16},
A.~M.~Koekemoer\altaffilmark{2},
P.~Kurczynski\altaffilmark{17},
E.~Merlin\altaffilmark{1},
A.~Mortlock\altaffilmark{18},
J.~A.~Newman\altaffilmark{19},
D.~Paris\altaffilmark{1},
L.~Pentericci\altaffilmark{1},
R.~Simons\altaffilmark{20},
S.~P.~Willner\altaffilmark{21}
  }
%\email{@}


\altaffiltext{1}{INAF -- Osservatorio Astronomico di Roma, via di
  Frascati 33, 00040  Monte Porzio Catone, Roma, Italy; \myemail}
\altaffiltext{2}{Space Telescope Science Institute, 3700 San Martin Drive, Baltimore, MD 21218, USA}
\altaffiltext{3}{Department of Physics and Astronomy, University of California, Riverside, CA 92521, USA}
\altaffiltext{4}{UCO/Lick Observatory and Department of Astronomy and
  Astrophysics, University of California, Santa Cruz, CA 95064 USA}
\altaffiltext{5}{Department of Astronomy, The University of Texas at Austin,  Austin, TX 78712, USA}
\altaffiltext{6}{Max-Planck-Institut f\"{u}r Extraterrestrische Physik (MPE), Postfach 1312, 85741 Garching, Germany}
\altaffiltext{7}{Department of Astronomy, University of Massachusetts, 710 North Pleasant Street, Amherst, MA 01003, USA}
\altaffiltext{8}{Center for the Exploration of the Origin of the Universe, Department of Physics and Astronomy,
Seoul National University, Seoul, Korea}
\altaffiltext{9}{National Optical Astronomy Observatories, 950 N Cherry  Avenue, Tucson, AZ 85719, USA}
\altaffiltext{10}{Joint ALMA Observatory, Alonso de Cordova 3107, Vitacura, Santiago, Chile}
\altaffiltext{11}{University of Nottingham, School of Physics and Astronomy, University of Nottingham,  Nottingham NG7 2RD, UK}
\altaffiltext{12}{Center for Galaxy Evolution, Department of Physics and
  Astronomy, University of California, Irvine, 4129 Frederick Reines
  Hall, Irvine, CA 92697, USA}
\altaffiltext{13}{Steward Observatory, University of Arizona, 933 North Cherry Avenue, Tucson, AZ 85721, USA}
\altaffiltext{14}{School of Physics and Astronomy, University of Nottingham, Nottingham, UK}
\altaffiltext{15}{Aix Marseille Universit\'{e}, CNRS, LAM (Laboratoire d'Astrophysique de Marseille) UMR 7326, 13388, Marseille, France}
\altaffiltext{16}{Department of Physics and Astronomy, University of
  Kentucky, Lexington, KY, USA}
\altaffiltext{17}{Department of Physics and Astronomy, Rutgers, The State University of New Jersey, Piscataway, NJ 08854, USA}
\altaffiltext{18}{SUPA Institute for Astronomy, University of  Edinburgh, Royal Observatory, Edinburgh EH9 3HJ, UK}
\altaffiltext{19}{Department of Physics and Astronomy, University of  Pittsburgh, Pittsburgh, PA 15260, USA}
\altaffiltext{20}{Department of Physics and Astronomy, Johns Hopkins University, 3400 N. Charles Street, Baltimore, MD
21218, USA}
\altaffiltext{21}{Harvard-Smithsonian Center for Astrophysics,
  Cambridge, MA 02138, USA}

%\altaffiltext{2}{CSIC}
%\altaffiltext{3}{CSIC}
%\altaffiltext{4}{CAUP}


\begin{abstract}
  We present the public release of the stellar mass catalogs for the
  GOODS-S and UDS fields obtained using some of the deepest near-IR
  images available, achieved as part of the Cosmic Assembly
  Near-infrared Deep Extragalactic Legacy Survey (CANDELS) project. We
  combine the effort from ten different teams, who computed the
  stellar masses using the same photometry and the same redshifts.
  Each team adopted their preferred fitting code, assumptions, priors,
  and parameter grid. The combination of results using the same
  underlying stellar isochrones reduces the systematics associated
  with the fitting code and other choices.  Thanks to the availability
  of different estimates, we can test the effect of some specific
  parameters and assumptions on the stellar mass estimate. The choice
  of the stellar isochrone library turns out to have the largest
  effect on the galaxy stellar mass estimates, resulting in the
  largest distributions around the median value (with a semi
  interquartile range larger than 0.1 dex).  On the other hand, for
  most galaxies, the stellar mass estimates are relatively insensitive
  to the different parameterizations of the star formation history.
  The inclusion of nebular emission in the model spectra does not have
  a significant impact for the majority of galaxies (less than a
  factor of 2 for $\sim$80\% of the sample). Nevertheless, the stellar
  mass for the subsample of young galaxies (age $<100$ Myr),
  especially in particular redshift ranges (e.g., $2.2< z<2.4$, $3.2<
  z<3.6$, and $5.5< z<6.5$), can be seriously overestimated (by up to
  a factor of 10 for $<20$ Myr sources) if nebular contribution is
  ignored.
\end{abstract}



\section{Introduction} \label{sec:intro}

Reliable stellar mass estimates are of crucial importance to achieve a
better understanding of galaxy evolution.  Stellar mass estimates are
complementary to other measures of galaxy stellar populations, such as
star formation rates (SFRs) and age. They tend to be more accurate
than estimates of SFR, which suffer from larger uncertainties due to
degeneracies between dust, age, and metallicity.

Nevertheless, stellar mass estimates are also potentially affected by
systematic uncertainties. The latter primarily originate from our
limited knowledge of several properties of the stellar populations,
such as their metallicity, which is not well constrained by a fit to
broad-band photometry \citep[e.g.,][]{castellano14}, the extinction
curve, or some phases of stellar evolution. The most striking example
is the thermally pulsating asymptotic giant branch (TP-AGB) phase
\citep{maraston05,marigo08} -- the modeling of which is still debated
\citep{zibetti13} -- which has a relevant contribution to the near-IR
emission of galaxies dominated by intermediate-age stellar populations
($\sim$1 Gyr). Another difficulty when estimating galaxy stellar
masses is properly reconstructing their star formation histories
(SFHs), which are usually approximated by simple (but not necessarily
appropriate) parametric functions.

Despite the systematics discussed above, high quality photometry and
accurate redshifts may significantly improve the reliability of the measured
stellar masses. The Cosmic Assembly Near-infrared Deep Extragalactic
Legacy Survey (CANDELS, \citealt{koekemoer11,grogin11}, PIs: S. Faber,
H. Ferguson) is of great help in this regard, thanks to its
exquisite quality near-IR photometry taken with the WFC3 camera on board the
Hubble Space Telescope (HST). CANDELS observations include some of the
deepest images in the visible and near-IR ever achieved over a wide
area, and have been complemented with the best auxiliary photometry
available in the mid-IR with Spitzer Space Telescope and in the ultraviolet with
  ground-based observations. Thanks to the combination of depth and area covered, 
stellar masses
from the CANDELS project can greatly improve our knowledge of the
galaxy stellar mass assembly process, both in a
statistical sense (e.g., they allow a robust measure of the galaxy
stellar mass functions; \citealt{grazian15}, G15 hereafter, \citealt{duncan14}) 
and for dedicated
analyses of interesting, faint and distant sources.

The aim of this paper is to present and accompany the release of
CANDELS stellar mass catalogs for the Great Observatories Origins Deep
Survey-South (GOODS-S, \citealt{giavalisco04}) and UKIRT Infrared Deep
Sky Survey (UKIDSS) Ultra-Deep Survey (UDS, \citealt{lawrence07})
fields.  This is the third of a series of papers that combine the
effort of several teams within the CANDELS collaboration to achieve an
improved result. In the first paper, \cite{dahlen13} (D13 hereafter)
presented and compared photometric redshifts computed by eleven teams
and demonstrated that the combination of multiple results is able to
reduce the scatter and outlier fraction in the photometric redshifts.
In the second work in the series \citep[][M15 hereafter]{mobasher14},
we performed a comprehensive study of stellar mass measurements and
analyzed the main sources of uncertainties and the associated error
budget by using mock galaxy catalogs based on semi-analytical models
as well as observed catalogs.  Biases of the ten different fitting
techniques turned out to be relatively small, and tended to be
confined to galaxies younger than $\sim$100 Myr, where models with a
fine spacing of the model grid in age and extinction appeared to
perform best. The faintest and lowest signal-to-noise ratio (S/N)
galaxies were found to be affected by the largest
scatter. Degeneracies between stellar mass, age, and extinction were
disentangled.

In this work we present and publicly release stellar masses computed
from the official CANDELS photometric and redshift catalogs by ten
teams, each adopting their preferred assumptions in terms of SFH,
stellar modeling and stellar parameters. We then combine these
estimates to suppress the effect of systematics deriving from the
choice of specific assumptions and priors in each of the methods.

The paper is organized as follows: Section~\ref{sec:data} presents
CANDELS photometric and redshift catalogs; Section \ref{sec:mass}
describes how stellar masses were estimated by the different teams,
whose results are compared in Section~\ref{sec:scattermethods}; the
official CANDELS stellar masses are presented in
Section~\ref{sec:refmass}; finally, we summarize the main results in
Section~\ref{sec:summary}. All magnitudes are in the AB system, and
the following cosmology has been adopted: H$_0$ = 70 km/s/Mpc,
$\Omega_{\small M} $ $=$ 0.3, and $\Omega_{\Lambda}$ = 0.7.

\section{Data set} \label{sec:data}

\subsection{GOODS-S and UDS multiwavelength catalogs}\label{sec:mwcat}

The CANDELS multiwavelength group has adopted a standardized method to
build catalogs in the CANDELS fields. Sources are extracted from the
CANDELS F160W mosaic using SExtractor. Total fluxes of the sources in
the high-resolution HST bands (WFC3 and ACS) are derived from the
aperture-corrected isophotal colors from SExtractor, run in dual mode
on PSF-matched images (where the PSF is the Point Spread Function). The photometry of the lower-resolution dataset
(e.g.,~ground-based and Spitzer) is derived using the template-fitting
software TFIT \citep{laidler07,papovich01}. In brief, TFIT uses the
{\it a-priori} information on the source location and surface
brightness profile on the F160W image to measure its photometry on the
low-resolution image. We refer the reader to \cite{galametz13} for
details on the adopted catalog building procedure.

The dataset available for each CANDELS field is rich but significantly
varies from field-to-field. As such, the multiwavelength catalogs of
GOODS-S and UDS contain different bands and data depths:

\paragraph{CANDELS-GOODS-S}
The GOODS-S catalog\footnote{Available at
  http://candels.ucolick.org/data\_access/GOODS-S.html} contains 34930
sources. The total area of $\sim 170$ square arcmin was observed by
WFC3 with a mixed strategy, combining CANDELS data in a deep (central
one-third of the field) and a wide (southern one-third) region with
ERS \citep{windhorst11} (northern one-third) and HUDF09
\citep{bouwens10} observations. The F160W mosaic reaches a 5$\sigma$
limiting magnitude (within an aperture of radius 0.17 arcsec) of 27.4,
28.2, and 29.7 in the CANDELS wide, deep, and HUDF regions,
respectively.  The multiwavelength catalog includes 18 bands: in
addition to the ERS/WFC3 and CANDELS/WFC3 data in the
$F105W$/$F125W$/$F140W$/$F160W$ filters, it also includes data from UV
($U$ band from both CTIO/MOSAIC and VLT/VIMOS), optical (HST/ACS
F435W, F606W, F775W, F814W, and F850LP), and infrared (HST/WFC3 F098M,
VLT/ISAAC $K_s$, VLT/HAWK-I $K_s$, and Spitzer/IRAC 3.6, 4.5, 5.8,
8.0$\mu$m) observations.  See \cite{guo13} for a summary of the
GOODS-S UV-to-mid-IR dataset and corresponding survey references.

\paragraph{CANDELS-UDS}
The UDS catalog\footnote{Available at
  http://candels.ucolick.org/data\_access/UDS.html} contains $35932$
sources distributed over an area of $\sim201.7$ square arcmin (roughly
a rectangular field of view of $22.3' \times 9'$). The F160W CANDELS
image reaches a 5$\sigma$ limiting depth of 27.45 within an aperture
of radius 0.20 arcsec. The multiwavelength catalog includes 19 bands:
the CANDELS data (WFC3 $F125W$/$F160W$ and ACS $F606W$/$F814W$ data),
$U$ band data from CFHT/Megacam, $B$, $V$, $R_c$, $i'$ and $z'$ band
data from Subaru/Suprime-Cam, $Y$ and $K_s$ band data from VLT/HAWK-I,
$J$, $H$ and $K$ bands data from UKIDSS (Data Release 8), and
Spitzer/IRAC data ($3.6$, $4.5$ from SEDS, \citealt{ashby13}, $5.8$
and $8.0\mu$m from SpUDS). The first released version of the catalog
contains a list of about $210$ sources with reliable spectroscopic
redshifts that we have extended in the present analysis with new
redshifts derived from the VLT VIMOS/FORS2 spectroscopic campaigns
(\citealt{bradshaw13}, \citealt{mclure13} and Almaini et al. in prep.)
and the MAGELLAN/IMACS spectroscopy presented in Appendix
\ref{app:magellan}.  See \cite{galametz13} for a summary of the UDS
UV-to-mid-IR dataset and corresponding survey references.

\subsection{Redshifts}\label{sec:redshifts}
 
CANDELS multiwavelength catalogs were cross-matched with a collection
of publicly available spectroscopic sources in both fields and with
the Magellan spectroscopy in UDS that is presented here for the first
time (see Appendix \ref{app:magellan}). $\sim 10\%$ and $\sim 2\%$ of
sources have a spectroscopic counterpart in GOODS-S and UDS,
respectively. In addition to the redshift value, the released catalogs
report the spectral quality and the original parent spectroscopic
survey. Once spectroscopic stars and poor quality spectra have been
removed, the fraction of galaxies with reliable spectroscopic
redshifts are $\sim 6\%$ and $\sim 1\%$ in the two fields,
respectively. If only H$_{160}<24$ sources are considered, the
spectroscopic fraction is $\sim 29\%$ and $\sim 8\%$, respectively.

Photometric redshifts have been computed for all CANDELS sources using
the official multiwavelength photometry catalogs described above.
Briefly, photometric redshifts are based on a hierarchical Bayesian
approach that combines the full PDF(z) distributions derived by
six\footnote{Five of the six photo-z methods include nebular lines.}
CANDELS photo-$z$ investigators. The 68\% and 95\% confidence
intervals, also included in the stellar mass catalogs, are calculated
from the final redshift probability distribution.  The techniques
adopted to derive the official CANDELS photometric redshifts, as well
as the individual values from the various participants, are described
in details by D13, and the photometric redshift catalogs of both
fields will be made available in a forthcoming paper (Dahlen et al. in
prep.).

\subsection{Data selection}\label{sec:selection}
 
We removed from the present analysis all objects flagged for having
bad photometry (see \citealt{guo13} and \citealt{galametz13}). This
information is available in the photometric catalogs as well in the
catalogs we release with the present work.

Stars have been classified either spectroscopically or
photometrically, and have been removed from the sample. 151 and 47
sources were identified as spectroscopic stars in GOODS-S and UDS,
respectively. No stellar mass nor any other parameter is provided in
the catalogs for these sources. Photometric stellar candidates have
been selected through the morphological information provided by
SExtractor on the F160W band (CLASS\_STAR$>$0.95) combined with the
requirement that S/N is larger than 20, which ensures reliability of
the CLASS\_STAR parameter.  The total number of high S/N point-like
candidates is 174 in GOODS-S and 224 in UDS.  Fainter point-like
sources are not flagged as stars due to unreliability of the
morphological criterion for low S/N sources, and conservatively
included in this analysis.

Finally, CANDELS catalogs have been cross-matched with X-ray sources
from the {\it Chandra} 4Ms catalogs of \cite{xue11} and \cite{rangel13}
\citep[see][]{hsu14} in the GOODS-S field and with the {\it XMM-Newton}
sample of \cite{ueda08} in the UDS field. We flag X-ray selected AGN
candidates and do not use them in the comparisons shown in this
paper. We remind the reader that a non detection in X-ray does not
prove that the source does not host an AGN, because AGN detection
depends on the depth of the X-ray survey and on the level of
obscuration. Dedicated works on IR AGN in all the CANDELS fields are
in preparation within the CANDELS collaboration.

Although we exclude AGNs from the present comparison, masses for AGN
candidates have been computed using the same technique as for non
active galaxies and are released in the catalogs.  In GOODS-S they
were computed by fixing the redshift to the photometric value obtained
by fitting the photometry with hybrid templates, as described by
\cite{hsu14}. This approach provides reliable mass estimates for the
large majority of obscured AGNs, whose SED is dominated by the stellar
component \citep{santini12b}. We caution that the stellar mass
estimate may not recover the true value for bright unobscured AGNs,
where ad-hoc techniques should be adopted
\citep{merloni10,santini12b,bongiorno12}. However, these sources make
up only 8\% of the AGN sample in GOODS-S \citep{hsu14}, which, due to
the small area, only rarely includes very bright AGNs. No dedicated
photometric redshifts for AGNs were computed in UDS. These are going
to be presented in a more complete future work on AGNs in all CANDELS
fields.

\section{Stellar mass estimates}\label{sec:mass}

\subsection{The estimate of the stellar mass}\label{sec:massest}

Stellar masses are commonly estimated by fitting the observed
multiwavelength photometry with stellar population synthesis templates
(e.g., \citealt{fioc97}, \citealt{bc03}, BC03 hereafter,
\citealt{maraston05}, \citealt{bruzual07b}, CB07 hereafter,
\citealt{conroy10}, see \citealt{conroy13} for a review).

The most widely used metric for goodness of fit is $\chi^2$.  A grid
for the free parameters must be set, and model spectra, computed by
fixing these parameters to the values in each step of the grid, are
compared with observed fluxes.  The best-fit parameters are either
provided by the template minimizing $\chi^2$ or computed as the median
of the Probability Distribution Function (PDF).  For some codes, the
PDF of stellar mass is computed from the $\chi^2$ contingency tables;
for others, the likelihood contours are determined in all the fitting
parameters using Markov Chain Monte Carlo (MCMC) sampling, and the
stellar mass PDF is determined by marginalizing over the other
parameters.  In this case, parameters are allowed to vary on a
continuous space, which is explored by means of a random walk.  In the
case of the SpeedyMC code \citep{acquaviva12}, used by one of the
participating teams, multi-linear interpolation between the
pre-computed spectra is used to compute the model SEDs. The final
best-fit parameters are the average of the posterior PDF, which is
proportional to the frequency of visited locations.

Free parameters in the models include stellar metallicity, age
(defined as time since the onset of star formation), dust reddening,
and the parameters describing the SFH of the galaxy.  Moreover,
several assumptions have to be made, such as the choice of the initial
stellar mass function (IMF) and the extinction law.

One of the most important assumptions for shaping the final templates
is the parameterization of the SFH. The star formation process in
galaxies can be very complicated and may have a stochastic nature. In
the attempt of reproducing the real SFH, several simple analytic
functions are usually adopted.  The most popular functions are:
\begin{itemize}
\item exponentially
declining laws, the so-called $\tau$ models (or direct-$\tau$ models):
$\psi(t)\propto exp(-t/\tau)$; 
\item exponentially increasing laws, also
called inverted-$\tau$ models: $\psi(t)\propto exp(t/\tau)$; 
\item constant
SFH: $\psi(t)=const$; 
\item instantaneous bursts: $\psi(t)\propto \delta(t_0)$. 
\end{itemize}
Some more complicated shapes include:
\begin{itemize}
\item  the
so-called delayed-$\tau$ models, i.e., rising-declining laws: e.g.,
$\psi(t)\propto t/\tau^2 \cdot exp(-t/\tau)$ or $\psi(t)\propto
t^2/\tau \cdot exp(-t/\tau)$;
\item  truncated SFH: $\psi(t)=const$ if
$t<t_0$, $\psi(t)=0$ otherwise;
\end{itemize}
and even more complex functional forms that have been proposed by previous
works \citep[e.g.,][]{behroozi13,simha14}. 

\subsection{Stellar mass estimates within CANDELS}
 
In order to investigate possible systematics due to different
assumptions, ten teams within the CANDELS collaboration computed the
stellar masses on the same released catalogs and on the same
redshifts. Good quality spectroscopic redshifts were used when
available, and the official CANDELS Bayesian photometric redshifts
(Dahlen et al. in prep.)  were adopted for all other galaxies.

Following the same notation as D13 and M15, we designate each team
with a code. Codes are composed of a number identifying the team PI,
of a letter indicating the stellar templates used, and if appropriate
the subscript $_\tau$ to indicate that purely exponentially decreasing
$\tau$ models have been assumed to parameterize the SFH.

Each of the teams was free to choose their favorite assumptions and
set their preferred parameter grid.  Although most of the teams
adopted BC03 stellar templates, other libraries were also used. Table
\ref{tab:param1} summarizes, for each of the participant teams, the
fitting technique, the code used to fit the data, the stellar
templates adopted, the main assumptions in terms of IMF, SFH,
extinction law, the ranges of the parameter grid employed, and the
priors applied to the template library. The grid steps differ one from
the other, as indicated in Table \ref{tab:param1}, and may in some
cases vary over the range covered.

Most mass estimates presented in this work adopt a \cite{calzetti00}
attenuation curve, while one method (Method 6a$_\tau$) treats the
extinction curve as a free parameter, and allows it to vary between a
\cite{calzetti00} and a SMC \citep{prevot84} one.

Three teams also include nebular emission lines and nebular continuum
in one case in addition to stellar emission in the model templates:
\begin{itemize}
\item Method 4b: the strength of the H$\beta$  line is computed from  the number of
ionizing photons for a given age and metallicity  assuming Case B
recombination and null escape fraction, and line ratios for 119 lines are taken from the Cloudy
models of \cite{inoue11} (see \citealt{salmon14}); 
\item  Method 11a$_\tau$: the addition of Ly$\alpha$, [OII], [OIII],
H$\alpha$ and H$\beta$ is done following the recipe of
\cite{ilbert09}, who adopted the 
relation from \cite{kennicutt98} between UV luminosity, SFR and [OII] flux, 
and applied line ratios to predict the flux in the other lines;
\item Method 14a: the flux from the nebular continuum and line
  emission is included by tracking the number of Lyman-continuum
  photons and assuming Case B recombination and null escape fraction,
  and modeling the empirical line intensities relative to H$\beta$ for
  H, He, C, N, O, S as a function of metallicity
  \citep{anders03,schaerer09,acquaviva11}.
\end{itemize}

Not all participants computed the stellar masses for both fields. In
the end, we present 9 different sets of stellar masses for the GOODS-S
field and 9 for the UDS field.
 
In addition to their preferred mass estimate, some of the teams also
provided further results based on different assumptions. Although we
will not use these to compute the final stellar mass values, we
present them in Table \ref{tab:param3} in Appendix
\ref{app:additional}, and we will use them to test how specific
parameters affect the best-fit result in the next section.

All these stellar mass estimates are included as electronic tables on
the online edition of this publication.

%-------------------------------------------------------------------------------------------------------

\begin{deluxetable*}{cccccc}
\tablecaption{\label{tab:param1} Summary of the assumptions adopted to
  compute the stellar masses in CANDELS.}
\tablehead{&\colhead{Method 2a$_\tau$} & \colhead{Method 2d$_\tau$} & \colhead{Method 4b } & \colhead{Method  6a$_\tau$} & \colhead{Method 10c}}
\tablecolumns{6}
\startdata
{\bf PI} & G. Barro & G. Barro & S. Finkelstein & A. Fontana & J. Pforr \\
\\[0.1pt]
{\bf fitting method} & min $\chi^2$ & min $\chi^2$ & min $\chi^2$ & min $\chi^2$ & min $\chi^2$ \\
\\[0.1pt]
{\bf code} & FAST\footnote{\cite{kriek09}.} v0.9b & Rainbow\footnote{\cite{perezgonzalez08}, \cite{barro11b}, https://rainbowx.fis.ucm.es/Rainbow\_Database/.} & own code & zphot\footnote{\cite{giallongo98}, \cite{fontana00}.} & HyperZ\footnote{\cite{bolzonella00}, http://webast.ast.obs-mip.fr/hyperz/.} \\
\\[0.1pt]
{\bf stellar templates} & BC03\footnote{\cite{bc03}.} & PEGASE\footnote{\cite{fioc97}.} v1.0 & CB07\footnote{\cite{bruzual07b}.} & BC03$^e$ & M05\footnote{\cite{maraston05}.} \\
\\[0.1pt]
{\bf  IMF} & Chabrier & Salpeter & Salpeter & Chabrier  & Chabrier \\
\\[0.1pt]
{\bf SFH} & $\tau$\footnote{Exponentially decreasing SFH (direct-$\tau$ models, see  Section~\ref{sec:massest}).} & $\tau$$^i$ & $\tau$$^i$ + inv-$\tau$\footnote{Exponentially increasing SFH (inverted-$\tau$ models, see  Section~\ref{sec:massest}).} &$\tau$$^i$ & $\tau$$^i$ + trunc.\footnote{Truncated SFH (see  Section~\ref{sec:massest}).}\\
  &   &   & + const.\footnote{Constant SFH (see  Section~\ref{sec:massest}).} &   & + const.$^l$ \\
\\[0.1pt]
{\bf log ($\tau$/yr)} & 8.5--10.0 & 6.0--11.0 & 5.0--11.0 & 8.0--10.2 & 8.0, 8.5, 9.0 \\
step\footnote{The number of steps is indicated when the grid size is not uniform over the range covered. }  & 0.2 & 0.1 & 6 steps & 9 steps &   \\
\\[0.1pt]
{\bf  log ($\tau^{INV}$/yr)\footnote{Timescale for inverted-$\tau$ models.}} &   &   & 8.5, 9.0, 10.0 &  &   \\
\\[0.1pt]
{\bf  log ($t_0$/yr)\footnote{$t_0$ is the timescale for truncated SFH (see Section~\ref{sec:massest}).}} &  &  &  &  & 8.0, 8.5, 9.0  \\
\\[0.1pt]
{\bf metallicity [Z$_\odot$] } &  1 & 0.005, 0.02, 0.2, &  0.02, 0.2, 0.4, 1 & 0.02, 0.2, 1, 2.5 &  0.2, 0.5, 1, 2.5  \\
  &   &  0.4, 1, 2.5, 5 &   &    &     \\
\\[0.1pt]
{\bf log (age/yr)} & 7.6--10.1 & 6.0--10.1 & 6.0--10.1 & 7.0--10.1 &8.0 -- 10.3 \\
step$^m$ & 0.1 & 60 steps & 40 steps & 110 steps & 221 steps \\
\\[0.1pt]
{\bf extinction law} & Calzetti & Calzetti & Calzetti & Calzetti + SMC &--- \\
\\[0.1pt]
{\bf extinction E(B-V)} &0.0--1.0 & 0.00--1.24& 0.0--0.8 & 0.0--1.1& 0.0 \\
step &0.025 &0.025& 0.02 & 0.05&  \\
\\[0.1pt]
{\bf nebular emission } & no &  no & yes  & no  &no  \\
\\[0.1pt]
 {\bf priors } &\footnote{Age must be lower than the age of the
   Universe at the galaxy redshift.} & $^p$& $^p$& $^p$ \footnote{Fit
   only fluxes at $\lambda_{RF}<5.5 \mu$m;
   $z_{form}\geq1/\sqrt{\tau}$, where $z_{form}$ is the redshift of
   the onset of the SFH; templates with E(B-V)$>$0.2 and
   age/$\tau$$>$3 or  with  E(B-V)$>$0.1 and Z/Z$_\odot$$<$0.1 or with
   age$>$1Gyr and  Z/Z$_\odot$$<$0.1 are   excluded.} & $^p$ \footnote{Fit only fluxes at $\lambda_{RF}<2.5 \mu$m.} \\
\\[0.1pt]
 {\bf reference } & 1 & 2 & 3 & 4 & 5 \\
\enddata
  \tablecomments{References. (1) \cite{barro13} ; (2)
    \cite{perezgonzalez08}; (3) \cite{finkelstein12}; (4)
    \cite{fontana06}; (5) \cite{daddi05,maraston06,pforr12,pforr13}.}
\end{deluxetable*}

%-------------------------------------------------------------------------------------------------------

\begin{deluxetable*}{cccccc}
\tablenum{1}
\tablecaption{\label{tab:param2} 
  ({\it continued})}
\tablehead{&\colhead{Method 11a$_\tau$} & \colhead{Method 12a } & \colhead{Method 13a$_\tau$} & \colhead{Method 14a} & \colhead{Method 15a}}
\tablecolumns{6}
\startdata
{\bf PI} & M. Salvato & T. Wiklind & S. Wuyts & B. Lee & S.-K. Lee\\
\\[0.1pt]
 {\bf fitting method} &  median of the & min $\chi^2$ & min $\chi^2$ & MCMC & min $\chi^2$ \\
 &  mass PDF\footnote[$s$]{Because Le Phare code does not compute the median mass when it is
   lower than $10^7M_\odot$, we use the minimum $\chi^2$ technique in
   these cases.} &  &  &   &   \\
\\[0.1pt]
{\bf code} & Le Phare\footnote[$t$]{Arnouts \& Ilbert, in preparation.} &  WikZ\footnote[$u$]{\cite{wiklind08}.} & FAST$^a$ v0.8b & SpeedyMC\footnote[$v$]{\cite{acquaviva12}.} & own code \\
\\[0.1pt]
{\bf stellar templates} & BC03$^e$ & BC03$^e$ & BC03$^e$ &  BC03$^e$ & BC03$^e$ \\
\\[0.1pt]
{\bf  IMF} &  Chabrier & Chabrier & Chabrier & Chabrier & Chabrier \\
\\[0.1pt]
{\bf SFH} & $\tau$$^i$ & del-$\tau$\footnote[$w$]{Delayed-$\tau$ models: $\psi(t)\propto t/\tau^2 \cdot exp(-t/\tau)$.}  & $\tau$$^i$  &  $\tau$$^i$  + del-$\tau$$^w$ +
const.$^l$ & del-$\tau$$^w$ \\
& & &  &  + lin. incr.\footnote[$x$]{Linearly increasing models: $\psi(t)\propto t$.}  & \\
\\[0.1pt]
{\bf log ($\tau$/yr)} & 8.0--10.5  & -$\infty$\footnote[$y$]{The
  $\tau$ grid starts from 0.0 Gyr in the linear space.} -- 9.3 &8.5--10.0 & 7.0--9.7 & 8.0--10.0\\
step$^m$  & 9 steps & 9 steps & 0.1 &  --- & 14 steps\\
\\[0.1pt]
{\bf metallicity [Z$_\odot$] } &  0.4, 1 & 0.2, 0.4, 1, 2.5 & 1 & 1& 0.2, 0.4, 1, 2.5 \\
\\[0.1pt]
{\bf log (age/yr)} & 7.0--10.1 & 7.7--9.8 & 7.7--10.1 & 8.0--10.1 &7.7--10.1\\
step$^m$ &  57 steps & 24 steps & 0.1 & --- & 64 steps\\
\\[0.1pt]
{\bf extinction law} & Calzetti & Calzetti & Calzetti & Calzetti & Calzetti \\
\\[0.1pt]
{\bf extinction E(B-V)} & 0.0--0.5& 0.0--1.0& 0.0--1.0 & 0.0--1.0 & 0.0--1.5\\
step & 0.1 & 0.025& 0.025 & --- & 0.025\\
\\[0.1pt]
{\bf nebular emission }  & yes  & no  & no & yes  & no \\
\\[0.1pt]
 {\bf priors } & $^p$ \footnote[$z$]{E(B-V) $<0.15$   if age/$\tau>4$.} &$^p$ & $^p$ & $^p$ & $^p$\\
\\[0.1pt]
 {\bf reference } & 6 & 7 & 8 & 9 & 10\\
\enddata
\tablecomments{References. (6) \cite{ilbert10}; (7)
  \cite{wiklind08,wiklind14}; (8) \cite{wuyts11b}; (9) B. Lee et
  al., in prep.; (10) \cite{lee10}.}
\end{deluxetable*}

%-------------------------------------------------------------------------------------------------------

\section{Comparison of different assumptions} \label{sec:scattermethods}

This section discusses the overall agreement/disagreement and then the
effects of the SFH and of including nebular emission.  For a more
detailed and systematic discussion about the uncertainties in the
stellar mass estimates of different methods, we refer the reader to
M15.

\subsection{Overall comparison} \label{sec:comp}

Because no ``true'' mass against which to compare all the others is
available, we start by comparing each mass estimate with the median
value. This median was computed for each galaxy by considering all
sets of stellar masses after rescaling them to the same Chabrier IMF
(as this is the IMF adopted by all but two of the teams): following
\cite{santini12a}, we subtract 0.24 dex from stellar masses computed
assuming the Salpeter IMF. To deal with the low number of measurements
available to compute the median, we adopt the Hodges-Lehmann
estimator, defined as the median value of the means in the linear
space of each pair of estimates in the sample:
\begin{equation}\label{eq:hlmean}
M_* = {\rm median} \left ( \frac{x_i+x_j}{2} \right ) ,
\end{equation}  
We used a bootstrap procedure (with 10 times the number of measurements
iterations) to randomly choose the pairs $i$ and $j$, rather than
using all possible values.  This statistical estimator has the
robustness of an ordinary median but smaller uncertainty.  We refer to
this Hodges-Lehmann mean value as $M_*^{\rm MEDIAN all}$.

\begin{figure*}[!t]
\resizebox{\hsize}{!}{
\includegraphics[angle=0]{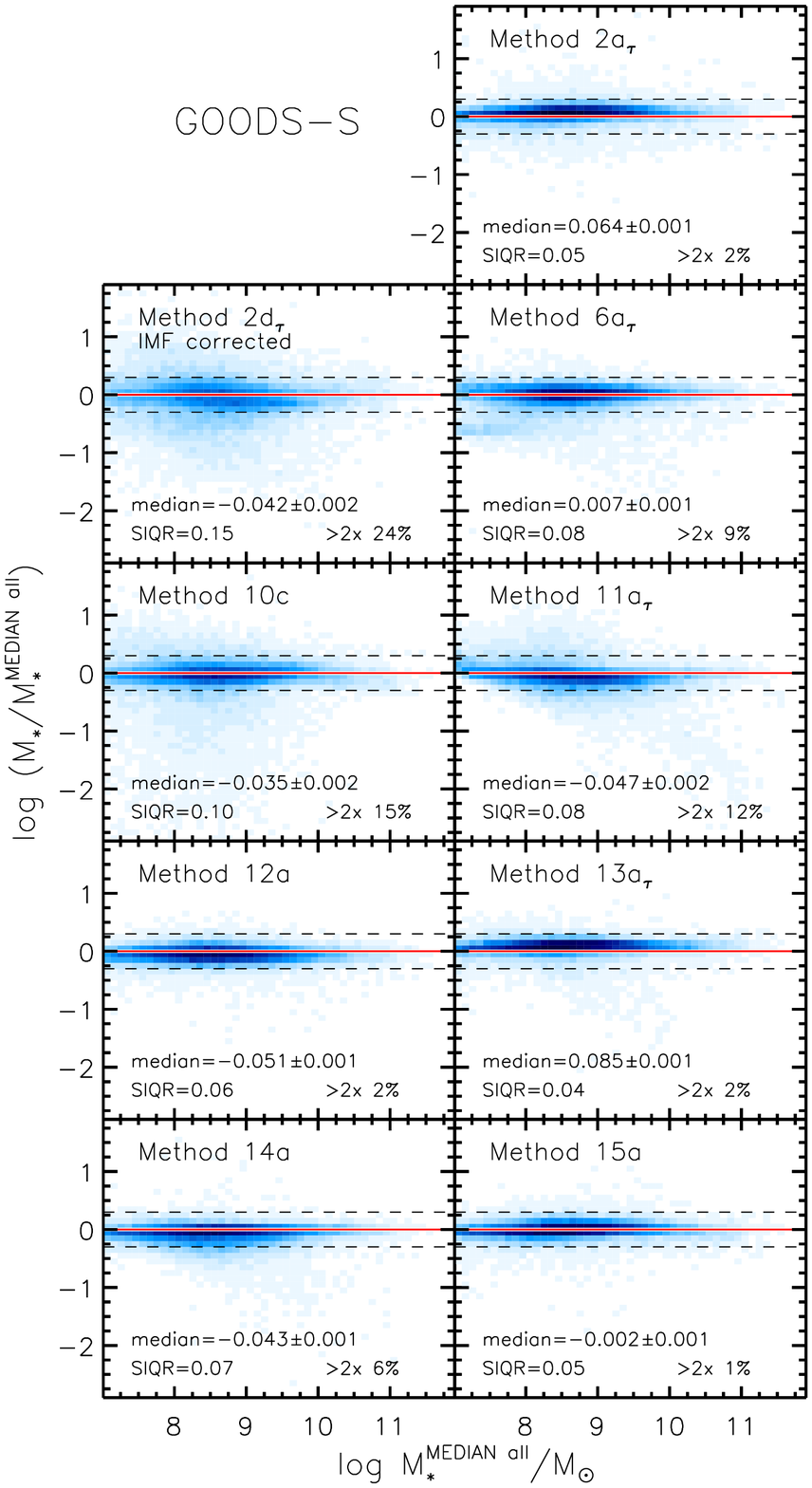}
\includegraphics[angle=0]{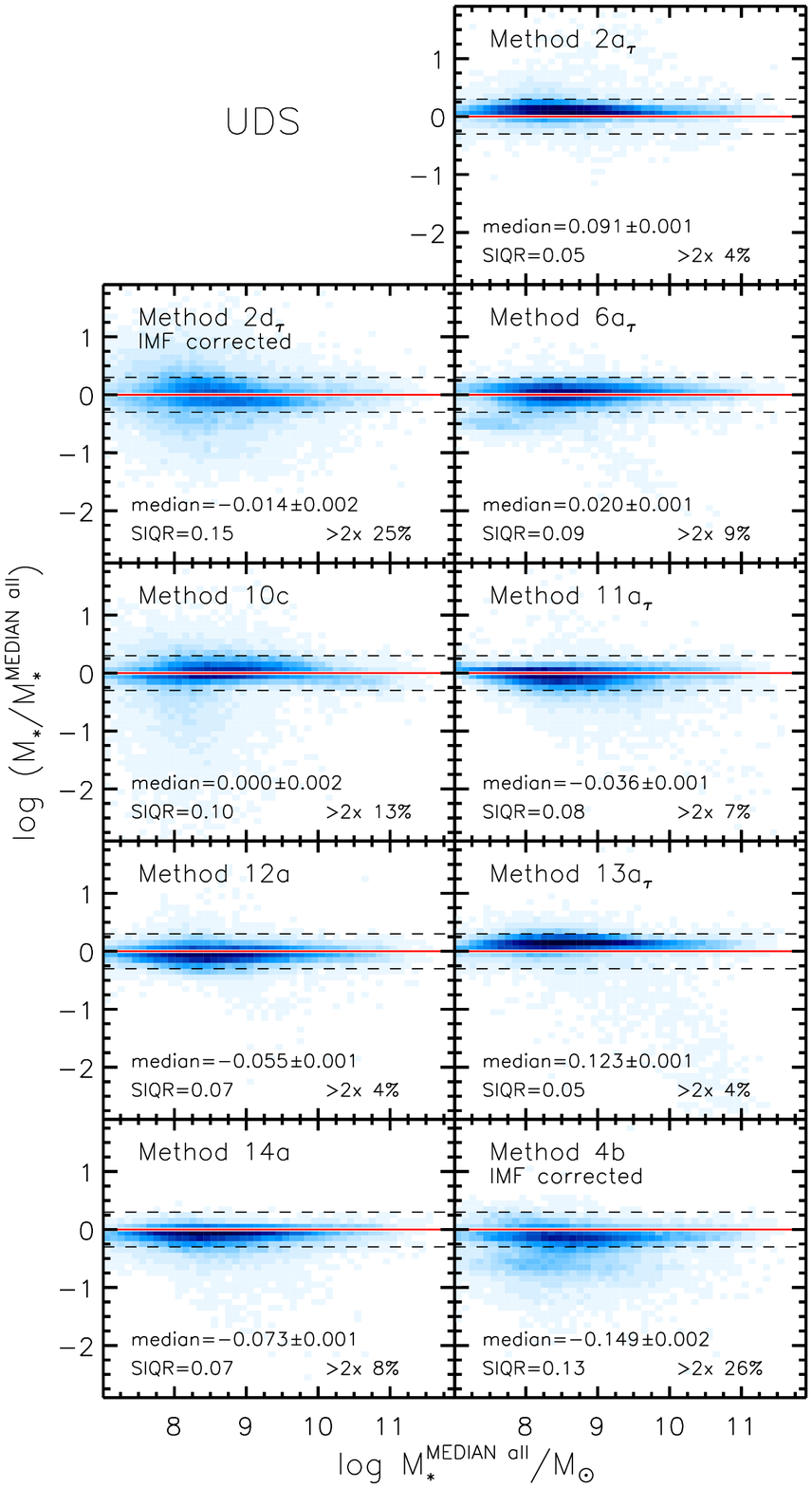}}
\caption{Comparison between the ratio of the different stellar mass
  estimates and the median mass for the GOODS-S ({\it left} panels)
  and the UDS ({\it right} panels) samples. All masses have been
  rescaled to the same Chabrier IMF. The plane is colored according to
  the density of sources, increasing from lightest to darkest shades
  on a linear scale. The red solid horizontal line indicate null
  difference with respect to the median mass, while the black dashed
  lines enclose the region where the difference is within a factor of
  2. The median logarithmic ratio with the associated error and the
  semi interquartile range (SIQR, see text) are printed in each panel,
  as well as the fraction of sources differing from the median value
  by more than a factor of 2 (i.e., falling beyond the dashed lines).}
\label{fig:cfr_allM}
\end{figure*}

Figure \ref{fig:cfr_allM} shows the ratio between each set of stellar
masses and the median value as a function of the median.  On average,
the agreement among the different estimates is quite satisfactory,
despite the different assumptions adopted.  The uncertainty associated
with the median value of the $\log(M_*/M_*^{\rm MEDIAN all})$ distribution
was computed as $\sigma/\sqrt{(2/\pi) N}$, where $\sigma$ is the
standard deviation and $N$ is the number of objects one takes the
median of, i.e., the number of galaxies in the sample. For most of the
stellar mass sets, the majority of values are tightly clustered around
the median of the distribution.  We quantify the broadness of the
distribution by means of the semi interquartile range (SIQR), defined
as half the difference between the $75^{th}$ and the $25^{th}$
percentile.  The typical SIQR is lower than 0.1 dex for most
estimates.  We also quantify the importance of the tails of the $\log
(M_*/M_*^{\rm MEDIAN all})$ distributions as the fraction of estimates
differing from the median value by more than a factor of 2.

The stellar mass estimates showing the largest deviations from the
median (SIQR$\sim$0.1-0.15) are those based on stellar templates other
than BC03 (Methods 2d$_\tau$, 4b and 10c). BC03 templates, adopted by
most of the teams, strongly constrain the median value. The same
methods also show the largest fraction of objects in the tails of the
distribution (13--26\%).  Two teams (Methods 4b and 10c) have used
stellar templates including a treatment of the TP-AGB phase
\citep{maraston05,marigo08}.  The enhanced emission at near-IR
wavelengths due to the contribution of TP-AGB stars, especially for
galaxies dominated by intermediate-age stellar populations ($\sim$1
Gyr), forces an overall lower normalization, hence slightly smaller
stellar masses (e.g., \citealt{maraston06,vanderwel06}; M15). However,
the scatter is larger than the offset (see also \citealt{santini12a}
for a comparison between BC03- and CB07-based results).

Methods 4b and 10c present other differences compared to the other
teams.  For example, Method 10c is the only one that does not include
extinction in the templates.  With the aim of verifying whether the
lack of extinction could be responsible for the large dispersion, we
estimated the stellar mass by using the very same assumption as Method
10c but allowing for dust extinction (see Method 10c$^{\rm dust}$
presented in Appendix \ref{app:additional}).  The SIQR and the
fraction of objects in the tails of the distribution are only very
mildly reduced if not unchanged (SIQR $=$ 0.09 and 0.08 and 15\% and
12\% of sources differ by more than a factor of 2 in GOOODS-S and UDS,
respectively). This implies that dust reddening alone cannot explain
the large scatter around the median value.

Both Methods 4b and 10c assume different SFH shapes instead of simple
$\tau$ models, adopted by 5 out of 10 teams. We will demonstrate in
Section~\ref{sec:sfh} that stellar masses are stable against different
parameterizations of the SFH (at least those we could directly test).
However, the tail of the $\log (M_*/M_*^{\rm MEDIAN all})$
distribution towards lower values disappears for Method 10c when only
galaxies best-fitted with direct-$\tau$ models are considered, and the
fraction of sources differing from the median by more than a factor of
2 decreases to 8\% and 11\% in GOODS-S and UDS,
respectively. Nevertheless, the SIQR is basically unchanged (0.09 and
0.11 dex in the two fields, respectively).

Method 4b shows a distribution with respect to the median value that
is slightly bimodal. This is partly responsible for its broadness.
This effect has already been reported by M15, who ascribe it to
parameter degeneracy combined with the adoption of different stellar
templates and SFH with respect to the majority of other mass
estimates. The degeneracy seems to cause the best solution to
fluctuate between either a population of lower mass galaxies (likely
young and star-forming) or alternatively a population of higher mass
ones, in agreement with the median values. Method 4b differs from most
of the others also due to the inclusion of nebular emission. Although
we will demonstrate in Section~\ref{sec:neb} that nebular emission
strongly affects stellar masses in only a subset of sources, it may
increase the degeneracy when combined with some peculiar SFH
parameterizations, e.g., inverted-$\tau$ models used by Method 4b.

Finally, Methods 2d$_\tau$ and 4b are based on a different choice of
the IMF. They were converted to a Chabrier IMF by multiplying the
masses by a constant value (i.e., by subtracting 0.24 dex, see above),
which is a good approximation. Indeed, the fact that the median value
of $\log (M_*/M_*^{\rm MEDIAN all})$ is close to zero for Method
2d$_\tau$ provides a further confirmation of the applicability of a
constant shift. For Method 4b the median is instead shifted to lower
values. This is likely a consequence of the effect of inclusion of
TP-AGB stars and parameter degeneracy discussed above. Anyway, the
bulk of the galaxies are located reasonably close to zero, once again
confirming the validity of the IMF conversion.

\subsection{Star formation histories} \label{sec:sfh}

To better investigate the effect of the assumed parameterization of
 the SFH, we take advantage of the additional mass estimates that were
provided by several teams and that are presented in Appendix
\ref{app:additional}. Because these results are based on the very same
assumptions as the mass estimates presented above except for the SFH
parameterization, they allow us to isolate its effect on the mass
estimates by leaving the other assumptions unchanged.

\begin{figure}[!t]
\resizebox{\hsize}{!}{\includegraphics[angle=0]{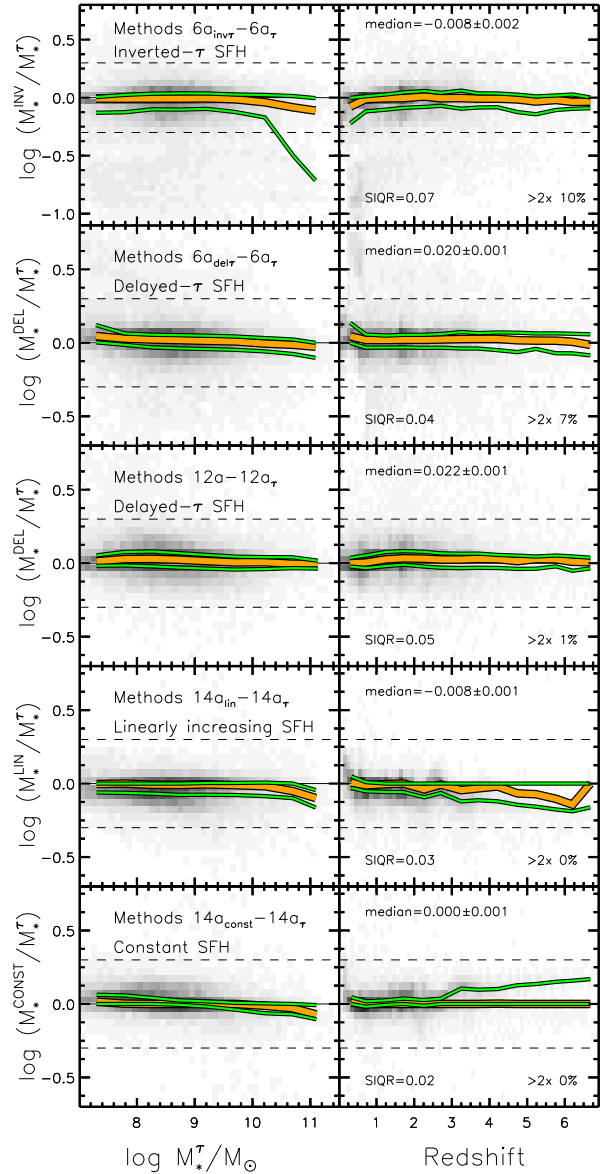}}
\caption{Comparison among stellar masses computed with consistent
  methods except for the choice of the SFH for the GOODS-S field. The
  ratio between masses computed with direct-$\tau$ models
  ($M_*^\tau$), inverted-$\tau$ models ($M_*^{INV}$), delayed-$\tau$
  models ($M_*^{DEL}$), linearly increasing models ($M_*^{LIN}$) and
  constant models ($M_*^{CONST}$) is studied against stellar mass
  ($M_*^\tau$, {\it left} panels) and redshift ({\it right}
  panels). Colors show density of sources as in
  Figure~\ref{fig:cfr_allM}. The thick orange lines show the median in
  bins of mass ($\Delta \log M_*=0.5$) or redshift ($\Delta z=0.5$),
  while the two thin green lines enclose 50\% of the sample
  (representing the 25$^th$ and 75$^th$ percentiles in the same
  bins). The median logarithmic ratio with the associated error and
  the semi interquartile range (SIQR, see text) are printed in each
  panel, as well as the fraction of sources differing from the median
  value by more than a factor of 2 (i.e., falling beyond the dashed
  lines). The black solid line shows the locus where the two stellar
  masses are comparable.  Delayed-$\tau$ models in Method 6a$_{\rm
    del\tau}$ have a slightly different analytic shape
  ($\psi(t)\propto t^2/\tau \cdot exp(-t/\tau)$) compared to Method
  12a ($\psi(t)\propto t/\tau^2 \cdot exp(-t/\tau)$).  }
\label{fig:sfh}
\end{figure}

Figure~\ref{fig:sfh} shows a comparison between the stellar masses in
the GOODS-S sample computed under the assumption of direct-$\tau$,
inverted-$\tau$, delayed-$\tau$ models, linearly increasing and
constant SFH. Stellar masses computed assuming such different SFH
parameterizations show a very good agreement with respect to
direct-$\tau$ models, with narrow distributions (SIQR$\leq$0.07 dex)
and no obvious offsets nor trends with stellar mass nor with redshift.
The only notable feature is a group of sources having $M_*^\tau$
(i.e., the mass based on direct-$\tau$ models) much larger (by up to
an order of magnitude) than $M_*^{INV}$ (i.e., the mass based on
inverted-$\tau$ models, upper panels of Figure~\ref{fig:sfh}): these
are galaxies which are old and massive according to the exponentially
decreasing model fit, and young and low-mass when fitted with an
exponentially increasing model. Indeed, sources with $\log
(M_*^{INV}/M_*^\tau)<-0.3$ have an average reddening E(B-V)$\sim$0.5
as inferred from the fit with inverted-$\tau$ models, but E(B-V)$<$0.1
when direct-$\tau$ models are adopted. On the other hand, sources
above this threshold show similar reddening in both fits. In any case,
such discrepant sources represent only 10\% of objects, the rest of
the sample showing $\log (M_*^{INV}/M_*^\tau)$ within a factor of 2.

Our results are in agreement with the previous work of \cite{lee10},
who concluded that stellar masses are robust and on average unaffected
by the choice of the SFH because of a combination of effects in the
estimate of the galaxy star formation rates and ages. However,
discrepant results were found by other groups
\citep{maraston10,pforr12}, who reported an underestimation of the
stellar mass when parameterizing the SFH as direct-$\tau$ models
compared to the mass predicted by semi-analytical models. According to
their analyses, the mismatch can be as much as 0.6 dex, with the exact
value depending on stellar mass, redshift, and fitting setup.

Our analysis does not include SFHs with bursts. However,
\cite{moustakas11} suggested that the adoption of smooth, dust-free
exponentially declining SFHs changes the stellar masses obtained with
bursty models (where bursts are added to direct-$\tau$ models) by no
more than 0.1 dex on average, and by less than a factor of 2 for
individual galaxies (see also \citealt{moustakas13} and references
therein).

\subsection{Nebular emission}\label{sec:neb}

\begin{figure*}[!t]
\resizebox{\hsize}{!}{\includegraphics[angle=90]{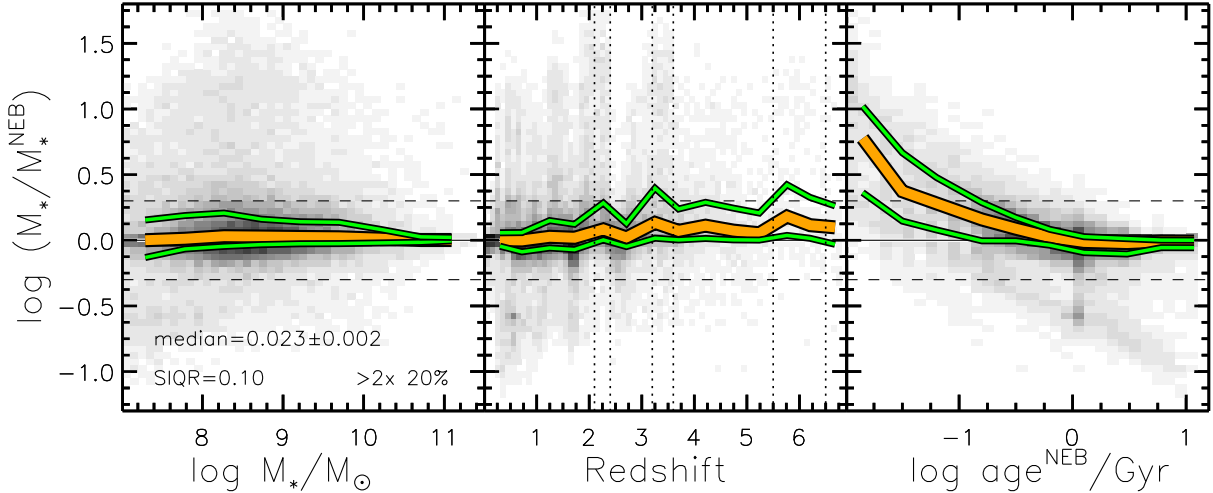}}
\caption{Ratio between masses estimated without ($M_*$) and with
  ($M_*^{\rm NEB}$) nebular emission as a function of $M_*$ ({\it
    left} panel), redshift ({\it central} panel), and age as inferred
  from the fit including nebular contribution (age$^{\rm NEB}$, {\it
    right} panel) for the GOODS-S field, using the estimates from
  Methods 6a$_\tau$ and 6a$_\tau^{\rm NEB}$. Colors and line styles
  are as in Figure~\ref{fig:sfh}.  Vertical dotted lines enclose three
  redshift ranges where strong nebular lines enter the near-IR
  filters, producing an overestimate of the stellar mass should they
  be ignored.}
\label{fig:neb}
\end{figure*}

Only three teams (Methods 4b, 11a$_\tau$ and 14a) have included
nebular emission in the stellar templates.  However, because these
three sets of stellar masses differ from the others also because of
other assumptions, it is not an easy task to isolate the effect of
nebular emission on the output stellar masses. For this reason, we
take advantage from the results of Method 6a$_\tau^{\rm NEB}$
presented in Appendix \ref{app:additional}, which differs from Method
6a$_\tau$ only due to the inclusion of nebular emission.

The comparison between stellar masses estimated without ($M_*$) and
with ($M_*^{\rm NEB}$) nebular emission is shown in
Figure~\ref{fig:neb}.  No significant offset is observed, nor are
there trends as a function of stellar mass or redshift.  Although the
$\log M_*/M_*^{\rm NEB}$ distribution is very wide, the semi
interquartile range is 0.1 dex, and 80\% of the sample is confined
within 0.3 dex from zero, meaning that the effect of nebular emission
on wide band photometry is weak for the bulk of the population (see
also G15). However, there are some exceptions that are worth
discussing here.

\begin{figure}[!h]
\resizebox{\hsize}{!}{\includegraphics[angle=0]{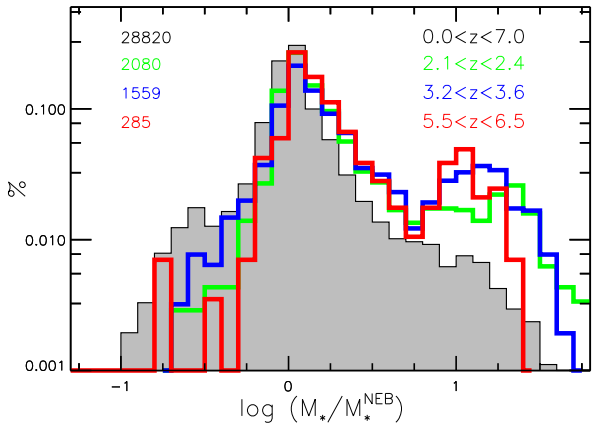}}
\caption{Normalized distribution of the ratio between masses estimated
  without ($M_*$) and with ($M_*^{\rm NEB}$) nebular emission in the
  entire sample (shaded histogram) and in the $2.1<z<2.4$ (green
  histogram), $3.2<z<3.6$ (blue) and $5.5<z<6.5$ (red) redshift
  ranges, for the GOODS-S sample and using the estimates from Methods
  6a$_\tau$ and 6a$_\tau^{\rm NEB}$. The numbers in the upper left
  corner show the number of galaxies in each sample, with the same
  color-code.}
\label{fig:histoneb}
\end{figure}

The effect of nebular emission is slightly enhanced in particular
redshift ranges, where strong nebular lines enter the near-IR filters,
on which stellar mass is strongly dependent. Moreover, this effect is
stronger at high redshift, where line emission contributes in a more
effective way to the light in broad-band filters due to wavelength
stretching as a consequence of cosmological expansion.  The three
redshift ranges where this effect is most evident are the $2.1<z<2.4$
region, where $J$, $H$ and $K$ band observe the [OII](3727\AA),
H$\beta$(4861\AA) and [OIII](5007\AA), and H$\alpha$(6563\AA) lines,
respectively, the $3.2<z<3.6$ region, where H$\beta$(4861\AA) and
[OIII](5007\AA) lines enter the $K$ band and [OII](3727\AA) line
enters the $H$ band, and the $5.5<z<6.5$ region, where
H$\beta$(4861\AA) and [OIII](5007\AA) lines are responsible for flux
enhancement in the 3.6$\mu$m band.  This is more clearly shown in
Figure~\ref{fig:histoneb}, where we plot the distribution of $\log
M_*/M_*^{\rm NEB}$ in the $2.1<z<2.4$, $3.2<z<3.6$, and $5.5<z<6.5$
redshift windows compared to the distribution of the total sample.
These redshift intervals show a positive tail in the distribution of
$\log M_*/M_*^{\rm NEB}$ due to misinterpretation of nebular line
emission as stellar continuum, resulting in a larger normalization in
the best-fit template, hence in a larger stellar mass estimate.  In
summary, the overestimate of the stellar mass when ignoring nebular
emission at high redshift, as predicted by previous works
\citep[e.g.,][]{stark13,schenker13}, is enhanced in particular
redshift ranges where strong lines enter the near-IR filters and does
not affect most of the galaxy sample.

\begin{figure}[!t]
\resizebox{\hsize}{!}{\includegraphics[angle=0]{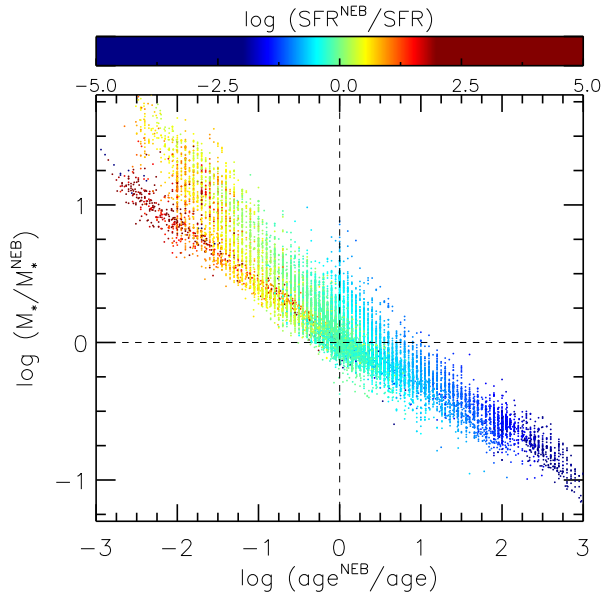}}
\caption{Ratio between masses estimated without ($M_*$) and with
  ($M_*^{\rm NEB}$) nebular emission as a function of the ratio
  between ages as inferred from the fit with (${\rm age}^{\rm NEB}$)
  and without (age) nebular emission, for the GOODS-S field and using
  the estimates from Methods 6a$_\tau$ and 6a$_\tau^{\rm
    NEB}$. Symbols are color-coded according to the ratio of their
  SFRs. }
\label{fig:dmdage}
\end{figure}

In addition, although $\log M_*/M_*^{\rm NEB}$ is on average
distributed around zero for the bulk of the population, its overall
distribution is asymmetric, with a tail towards higher values (see
shaded histogram in Figure~\ref{fig:histoneb}).  Because the spectra
of young and star-forming galaxies are characterized by many (and
intense) nebular emission lines, we expect that the effect of
including or neglecting nebular emission may depend on the galaxy
age. In fact, a well defined trend of $\log M_*/M_*^{\rm NEB}$ can be
observed as a function of age as inferred from the fit including the
nebular contribution (age$^{\rm NEB}$, right panel of
Figure~\ref{fig:neb}): stellar masses turn out to be severely
overestimated (by as much as an order of magnitude) in young
(age$^{\rm NEB}$$<$100 Myr) galaxies if nebular emission is
ignored. Indeed, strong nebular lines in their spectra may mimic the
Balmer and 4000\AA\ breaks, resulting in these sources being fitted
better by older, passive, and more massive templates \citep[see
also][]{atek11}.  To further explore this effect, we plot in
Figure~\ref{fig:dmdage} the ratio between stellar masses ($\log
M_*/M_*^{\rm NEB}$) as a function of the ratio of ages ($\log {\rm
  age}^{\rm NEB}/{\rm age}$), and color-code symbols according to the
ratio of their SFRs ($\log {\rm SFR}^{\rm NEB}/{\rm SFR}$), where the
superscript NEB denotes the parameters inferred from the fit with
templates including nebular emission. The stellar mass ratio is
strongly correlated with the age ratio as well as with the SFR ratio.
As noted above, ignoring nebular emission produces an overestimate of
the stellar mass in galaxies that are young and star-forming according
to the fit with nebular emission, while the stellar mass is mostly
unaffected when the best-fit age and SFR computed with and without the
nebular contribution are consistent. In general, when comparing two
distinct estimates of the stellar mass, the largest differences are
observed in galaxies that have a young star-forming solution in one
fit and an old passive solution in the other, as shown in
Figure~\ref{fig:dmdage}.  This behavior is common to every pair of
fits and not specific to the case with/without nebular emission, as
also discussed by M15.

Finally, in order to understand how important is the particular
implementation of nebular emission compared to including it at all, we
compared the results of the four methods that include the nebular
component, i.e., Methods 4b, 6a$_\tau^{\rm NEB}$, 11a$_\tau$ and
14a. More specifically, we compared each method with the median of the
four, in a similar way as done in Figure \ref{fig:cfr_allM}, after
rescaling all masses to the same Chabrier IMF.  The methods accounting
for nebular emission generally agree with each other better than they
do with the methods not including it: the distributions of the ratio
between each of the methods and their median has a SIQR of 0.05--0.1
dex, with a fraction of $<15\%$ of sources differing by more than a
factor of 2. The only exception is Method 4b. However, we ascribe its
wider distribution to the different stellar isochrone models adopted
by this method, as discussed in Section \ref{sec:comp}.

To summarize, Figures \ref{fig:neb}, \ref{fig:histoneb}, and
\ref{fig:dmdage} show that 80\% of the population is unaffected by the
inclusion of nebular emission (see also G15), and $\log M_*/M_*^{\rm
  NEB}$ is consistent with 0. For these galaxies, the inclusion of
nebular emission does not change the stellar mass by more than a
factor of 2.  However, nebular emission can strongly affect the light
emitted by subclasses of galaxies, notably galaxies in particular
redshift ranges (especially at $z>3$) or young (${\rm age}^{\rm
  NEB}<100$Myr) sources. For extremely young (${\rm age}^{\rm
  NEB}\lesssim20$Myr) galaxies, ignoring nebular emission may produce
an overestimate of the stellar mass by more than a factor of 10. The
sources whose difference in the stellar mass is larger than a factor
of 6 (i.e., beyond the minimum shown by the histograms in Figure
\ref{fig:histoneb}) is 6\% of the total sample.

\section{CANDELS reference stellar masses} \label{sec:refmass}

\subsection{The median mass approach}\label{sec:median}

Given the results shown in the previous sections, we can conclude that
the stellar mass is a stable parameter against different assumptions
in the fit for the majority of the galaxy population. Except for the
IMF, which introduces a roughly constant offset, the most important
assumption is the choice of the stellar population synthesis templates
(see also M15), which seem to severely affect stellar mass
estimates. Because BC03 templates are assumed by most of the teams (7
out of 9 for the GOODS-S sample, 6 out of 9 for the UDS sample), we
decided to compute a median mass by only considering BC03-based
estimates: in the GOODS-S field we consider stellar masses from
Methods 2a$_\tau$, 6a$_\tau$, 11a$_\tau$, 12a, 13a$_\tau$, 14a, 15a,
while in the UDS field we consider masses from Methods 2a$_\tau$,
6a$_\tau$, 11a$_\tau$, 12a, 13a$_\tau$, 14a.  As demonstrated by M15,
the median value provides the most accurate measure of the stellar
mass, as long as the individual estimates are unbiased compared to
each other, as is the case (see previous Section).

Because only a limited number of measurements are available, we
adopted the Hodges-Lehmann estimator (see Equation \ref{eq:hlmean})
instead of the median, as explained in Section~\ref{sec:comp},
computed in linear space.  We refer to this median value as $M_*^{\rm
\rm  MEDIAN}$ and consider it the reference mass for the CANDELS
catalogs.

We quantify the scatter around the median value caused by the
different assumptions for computing stellar masses by means of the
standard deviation ($\sigma_{M,{\rm CANDELS}}$) of the various
methods, again computed in linear space and only considering the
methods adopting BC03 stellar templates.  Therefore, this scatter does
not, by definition, account for the effect of stellar evolution
modeling, such as for example the inclusion of the TP-AGB phase, but
it is mainly caused by differences in the technicalities in the mass
computation, in the parameter grid sampling, in the assumed SFH, and
in the prior assumptions. This scatter is roughly 25-35\% of the
median values and shows no trend with age nor with rest-frame colors.

\begin{figure*}[!t]
\resizebox{\hsize}{!}{\includegraphics[angle=90]{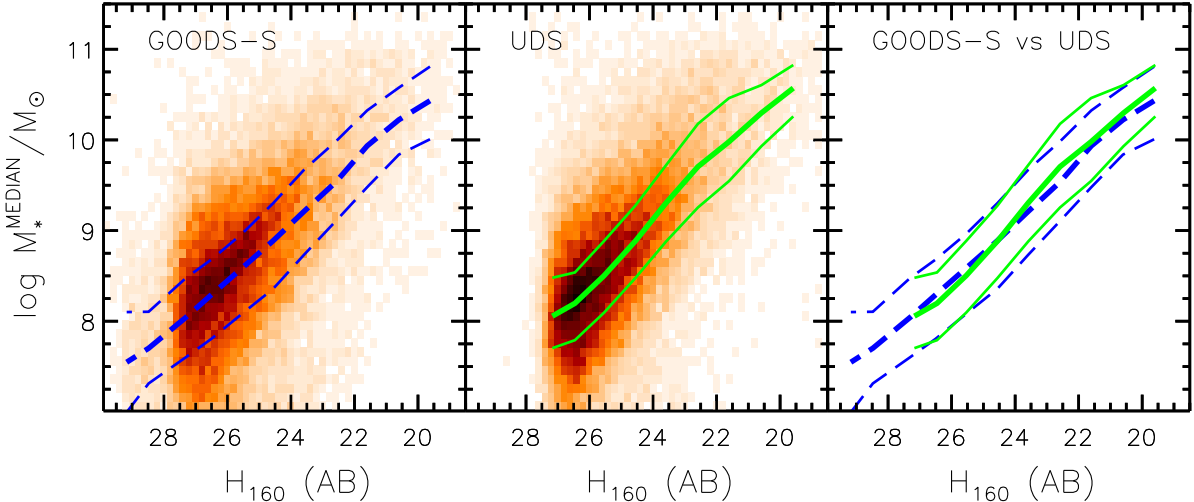}}
\caption{Reference median stellar mass (see Section \ref{sec:median})
  as a function of the observed $H_{160}$ band magnitude in GOODS-S
  ({\it left} panel) and UDS ({\it central} panel). The color reflects
  the density of sources, increasing from lightest to darkest on a
  linear scale. Thick lines show the median mass in bins of one
  magnitude, while thin lines show the semi interquartile range in the
  same bins. Medians and semi interquartile ranges are also reported
  in the {\it right} panel for a direct comparison between the two
  fields (GOODS-S: blue dashed; UDS: green solid).}
\label{fig:hmass}
\end{figure*}

\begin{figure*}[!t]
\resizebox{\hsize}{!}{\includegraphics[angle=90]{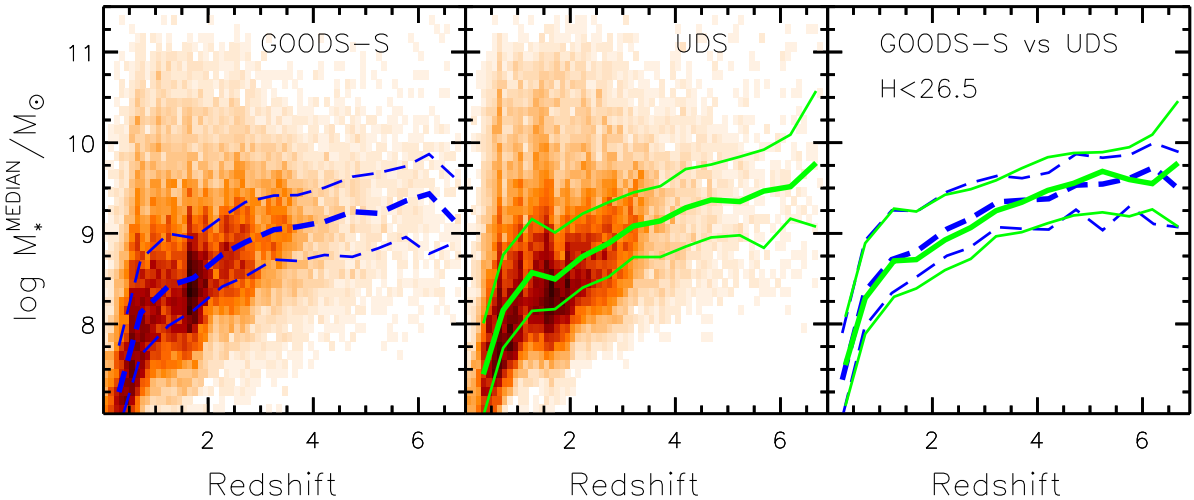}}
\caption{Reference median stellar mass (see Section \ref{sec:median})
  as a function of redshift in GOODS-S ({\it left} panel) and UDS
  ({\it central} panel). Colors and line styles are as in
  Figure~\ref{fig:hmass}. Thick lines show the median mass in bins of
  redshift ($\Delta z=0.5$), while thin lines show the semi
  interquartile range in the same bins. The {\it right} panel shows
  medians and semi interquartile ranges for the two samples cut at
  $H_{160}<26.5$. }
\label{fig:zmass}
\end{figure*}

Figure~\ref{fig:hmass} shows CANDELS reference stellar masses as a
function of observed $H_{160}$ band magnitude, while
Figure~\ref{fig:zmass} shows them as a function of redshift, in
GOODS-S and UDS.  The right panels of both figures show a comparison
between the two fields. In Figure~\ref{fig:zmass} we cut both samples
to $H_{160}<26.5$ to account for different observational depth in the
two fields. The $M_*$--$H_{160}$ and $M_*$--$z$ relations agree very
well in the two fields, confirming the consistency of the photometry,
i.e., meaning that there are no issues associated with colors or
photometric redshifts.

\subsection{Overall scatter among different methods vs scatter due to model degeneracy and photo-$z$ uncertainty}

All the stellar mass estimates that we are considering and comparing
in the present work are computed by assuming the photometric or
spectroscopic redshift at their reported value, not accounting any
redshift uncertainties. However, as shown by G15 when measuring galaxy
stellar mass functions, the reliability of photometric redshift may
represent the major source of uncertainty in the stellar mass
estimate. Moreover, the effect of model degeneracy (i.e., the
possibility of having two or more different SEDs fitting the
observations equally well because of the similar effects on broad-band
colors due to varying SFH, age, metallicity, dust) may be significant
and should not be underestimated.  This effect is reflected by the
width of the probability distribution function PDF($M_*$).

We estimate here the relative uncertainty on the stellar mass
originating from the scatter in the photometric redshifts combined
with model degeneracy and compare it with the systematic uncertainty
caused by adoption of different assumptions (but the same redshifts)
from the various teams.  We quantify relative uncertainties as
$\delta_{\rm M} = \sigma/M_*$, where $\sigma$ is the standard
deviation of the mass distribution for each object.

The relative uncertainty in the stellar mass estimates caused by
different assumptions in the fit is indicated as $\delta_{M,{\rm
    CANDELS}}=\sigma_{M,{\rm CANDELS}}/M_*^{\rm MEDIAN}$. For each
object, $\sigma_{M,{\rm CANDELS}}$ was calculated as explained in the
previous Section.

The relative uncertainty derived from model degeneracy and
uncertainties in the photometric redshift determination, denoted by
$\delta_{M,z}=\sigma_{M,z}/M_{* M,z}$, depends on the width of the
redshift probability distributions function PDF($z$) (D13) and on
photometric errors. It was estimated by means of a Monte Carlo
simulation as explained by G15.  Briefly, for each galaxy lacking a
spectroscopic estimate, a random redshift was extracted according to
its PDF($z$), and a PDF($M_*$) was computed by fitting the observed
photometry at the extracted redshift following Method 6a$_\tau$. For
spectroscopic sources, a single PDF($M_*$) was calculated by fixing
the redshift to its spectroscopic value. A mass estimate was then
extracted according to the PDF($M_*$).  This procedure was repeated
10000 times for each object and a standard deviation ($\sigma_{M,z}$)
and a mean value ($M_{* M,z}$) of the resulting mass distribution were
computed in linear space.  $\delta_{M,z}$ is included in the released
mass catalogs.

\begin{figure}[!t]
\resizebox{\hsize}{!}{\includegraphics[angle=0]{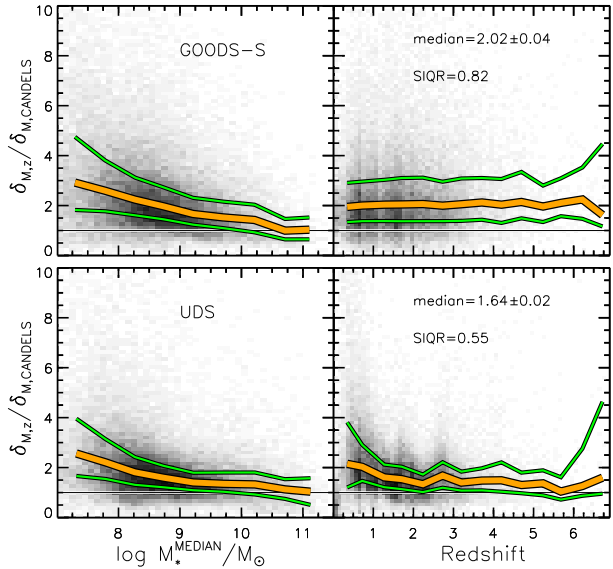}}
\caption{Ratio between the relative uncertainty caused by model
  degeneracy and scatter in the photometric redshifts ($\delta_{M,z}$)
  and that due to the adoption of different assumptions in the SED
  fitting ($\delta_{M,{\rm CANDELS}}$), as a function of the reference
  median mass ({\it left} panels) and redshift ({\it right}
  panels). {\it Upper} and {\it lower} panels show the GOODS-S and UDS
  samples, respectively. Colors and line styles are as in
  Figure~\ref{fig:sfh}. The black horizontal line shows the locus
  where the two uncertainties are comparable.}
\label{fig:scatter}
\end{figure}

Figure \ref{fig:scatter} shows the ratio between the relative
uncertainty due to model degeneracy and photometric redshifts scatter
($\delta_{M,z}$) and that due to differing assumptions in the SED
fitting ($\delta_{M,{\rm CANDELS}}$), as a function of stellar mass
and redshift.  $\delta_{M,z}$ is on average larger than
$\delta_{M,{\rm CANDELS}}$, at least for the bulk of the population,
by a factor of $\sim$2.  This factor increases for low-mass sources
and is close to unity at high stellar masses. Indeed, massive galaxies
have more accurate photometry thanks to the higher S/N, hence have
more accurate photometric redshifts and allow a lower level of model
degeneracy in the fit. No trend is observed with redshift.

The fact that the uncertainty due to the assumptions in the fit is on
average smaller than that originating from models degeneracy and
photo-$z$ scatter further justifies our choice of computing the median
of estimates performed by the various teams, despite the different
assumptions (see Table \ref{tab:param1}).

It is interesting to compare the value of $\delta_{M,z}$ in
spectroscopic and photometric sources, in order to have an idea of the
contribution of photometric redshift scatter compared to model
degeneracy. The average $\delta_{M,z}$ is about two times larger for
sources lacking good quality spectra compared to spectroscopic
galaxies.  Photometric redshifts scatter therefore makes the
uncertainty in stellar masses worse by a factor of 2, compared to what
inevitably one gets due to model degeneracy. However, the
spectroscopic sample is biased toward the brightest galaxies, which
have the cleanest photometry and therefore suffer from the lowest
level of model degeneracy in the fit. Fainter galaxies are susceptible
to a higher degeneracy.

In conclusion, Figure~\ref{fig:scatter} illustrates that model
degeneracy and photometric redshift scatter remain the largest source
of uncertainty when estimating stellar masses: the relative mass
uncertainty due to model degeneracy and photo-$z$ exceeds that due to
other systematics (assumptions and details in the fits and choice of
the parameters grid) by a factor of 2 for the bulk of the population,
and potentially by factors of ~several toward lower masses. Only for
the most massive galaxies is the contribution of model degeneracy
associated with photo-$z$ scatter comparable to that of systematics in
the mass computation.

\subsection{Comparison with 3D-HST stellar masses} \label{sec:3dhst}

\begin{figure}[!t]
\resizebox{\hsize}{!}{\includegraphics[angle=0]{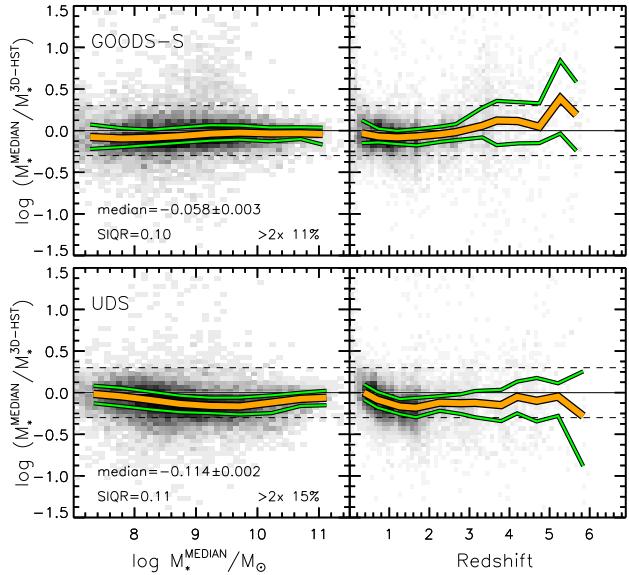}}
\caption{Ratio between CANDELS median masses and 3D-HST masses as a
  function of CANDELS median masses ({\it left} panels) and CANDELS
  redshifts ({\it right} panels) for the GOODS-S field ({\it upper}
  panels) and the UDS fields ({\it lower} panels). Only sources whose
  redshifts differ by less than 0.1 have been considered. Colors and
  line styles are as in Figure~\ref{fig:sfh}.  }
\label{fig:3dhst}
\end{figure}

We compare CANDELS reference stellar masses with those released by the
3D-HST team for the same fields \citep{skelton14}, by matching the two
catalogs in position with a tolerance of 0.1 arcsec.  They use BC03
templates, assume a Chabrier IMF, direct-$\tau$ model SFHs with a
minimum timescale $\log(\tau/yr)$ of 7, Solar metallicity, a minimum
$\log({\rm age}/yr)=7.6$, a Calzetti extinction law with
0.0$<$E(B-V)$<$1.0, and do not include nebular emission.  Because
their masses are computed on different photometric catalogs and hence
assuming different photometric redshifts, we only consider those
sources whose redshifts differ by less than 0.1.  The requirement that
positions and redshifts are within the tolerance leaves us with
$\sim40\%$ of the CANDELS sample.  The comparison, as a function of
both CANDELS median masses and redshifts, is shown in
Figure~\ref{fig:3dhst}.

The comparison is satisfying at a first-order glance, with quite
narrow dispersion (SIQR $\sim 0.1$ dex), increasing towards high
redshifts, where sources are generally fainter and photometry is
noisier.  However, a slight negative median offset ($\sim$$-0.1$ dex)
in both fields and a curved trend in UDS are observed.  These effects
could be ascribed to systematics in the SED fitting, in particular
affecting 3D-HST results. Indeed, while 3D-HST stellar masses result
from a single fit, CANDELS stellar masses are computed with a median
approach, which is able to wash out systematic biases affecting
specific assumptions in the SED modelling and was demonstrated to
provide a more robust measure (M15). To check whether this may be the
cause, we compare 3D-HST masses with the mass estimate within CANDELS
whose method and assumptions are closest 3D-HST ones, i.e. Method
2a$_\tau$.  The offset is reduced (median $=0.010 \pm 0.003$ in
GOODS-S and median $=-0.060 \pm 0.002$ in UDS), but the curved trend
is mostly unchanged. The residual offset observed in the UDS field and
the curved trend may be due to systematics in the photometry. Indeed,
UDS, having data of poorer quality, might be more susceptible to
biases compared to the better quality data in GOODS-S.

The difference between CANDELS and 3D-HST stellar mass estimates (SIQR
$\sim 0.1$ dex) is only slightly larger than the typical dispersion
observed among different estimates within CANDELS assuming the same
stellar models. However, the distributions are overall more extended:
the fraction of objects differing by more than a factor of 2 is 11\%
and 15\% in GOODS-S and UDS, respectively. If the distributions are
shifted so that the average offset is zero, these fractions are
marginally reduced (10\%). They are equal to 12\% and 11\% if 3D-HST
masses are compared to Method 2a$_\tau$ instead of the median CANDELS
masses, reflecting an intrinsic larger distribution due to differences
in the photometry and in the redshifts in addition to the systematics
in the mass estimates.

\subsection{CANDELS stellar mass catalogs} \label{sec:catalogs}

The median approach has revealed a powerful tool to overcome
systematics associated with the choice of the SED fitting parameters
(see previous section and M15). For this reason, the catalogs
presented here include the CANDELS reference median mass
($M_*^{\rm MEDIAN}$) as well as the median of only the methods including
nebular emission (and based on the same stellar isochrone library,
i.e., Methods 6a$_\tau^{NEB}$, 11a$_\tau$ and 14a), in both cases with
their associated scatters (see Section \ref{sec:median}).
Nevertheless, the catalogs also contain each individual estimate of
the stellar mass for each source (Tables \ref{tab:param1} and
\ref{tab:param3}). The various sets of stellar mass measurements agree
on average, at least those based on the same stellar
templates. However, the different assumptions, such as the SFH
parameterization or the inclusion of nebular emission, may produce
very different results for few peculiar and interesting objects or for
subsets of very young galaxies (see for example
Figure~\ref{fig:neb}). The comparison of the various methods may then
provide interesting information and a deeper insight in the galaxy
evolution paradigm.

In addition to the stellar masses, we also include in a separate file
other physical parameters (such as age, SFR, metallicity, dust
reddening, etc) and rest-frame magnitudes as estimated from the
various methods.  The same file also incorporates the uncertainties in
the stellar masses, neglecting redshift errors, associated with
methods 6a$_\tau$, 11a$_\tau$ and 12a.

Both files for each field are released as electronic tables in the
online version of the Journal and are uploaded on the STScI MAST
archive website for
CANDELS\footnote{http://archive.stsci.edu/prepds/candels/}.  
Tables showing the list of columns in each catalog are reported in
Appendix \ref{app:masscat}. The
catalogs are also available in the Rainbow
Database\footnote{http://arcoiris.ucolick.org/Rainbow\_navigator\_public,
  http://rainbowx.fis.ucm.es/Rainbow\_navigator\_public}
\citep{perezgonzalez08,barro11a}, which features a query menu that
allows users to search for individual galaxies, create subsets of the
complete sample based on different criteria, and inspect cutouts of
the galaxies in any of the available bands. It also includes a
crossmatching tool to compare against user uploaded catalogs.

\section{Summary} \label{sec:summary}

This paper accompanies the public release of the CANDELS mass catalogs
for the GOODS-S and UDS fields. We present and make publicly available
the reference CANDELS stellar masses obtained by combining the results
from various teams within the collaboration. Masses are estimated
adopting the official CANDELS photometry and assuming spectroscopic
redshifts when available, or the official CANDELS photometric
redshifts (Dahlen et al. in prep.) otherwise.  We also release the
individual stellar mass estimates computed by each team, as their
comparison may be useful especially when studying peculiar objects, as
well as other physical parameters (such as age, SFR, metallicity, dust
reddening, etc) and rest-frame magnitudes associated with the SED
fitting.

The availability of several mass estimates has allowed us to compare
methods, from which we conclude the following:

\begin{itemize}

\item The results from the various teams are in overall good agreement
  despite the different methods, assumptions, and priors.

\item The parameter which has the greatest effect on the stellar mass
  is the stellar isochrone library: due to different modeling of
  several stellar evolutionary phases, such as the TP-AGB phase, the
  adoption of different libraries produces a dispersion (in terms of
  the semi interquartile range) larger than 0.1 dex, while the
  dispersion shown by estimates based on the same stellar templates is
  on average smaller.

\item Stellar masses are stable against the choice of the SFH
  parameterization and differences in the metallicity/extinction/age
  parameter grid sampling.

\item The IMF only affects stellar masses as an overall scaling, as
  expected.

\item The inclusion of nebular emission can have a large effect but
  only on a small fraction of the sample. Specifically, ignoring
  nebular emission in the model spectra may cause an overestimate of
  the stellar mass in galaxies that are either young ($<100$Myr) or
  lie in particular redshift ranges, especially at high
  redshift. Indeed, in these redshift windows strong nebular lines
  enter the near-IR filters, on which stellar mass is strongly
  dependent. The overestimate can exceed a factor of 10 in extremely
  young ($<20$Myr) galaxies. Nevertheless, the effect of nebular
  emission is negligible for the majority (80\%) of galaxies. As
  current and future surveys push observations towards the youngest
  Universe, it becomes crucial to investigate whether the assumptions
  adopted to include nebular emission in the models are correct, in
  order to infer the most reliable stellar masses for the first
  galaxies.

\end{itemize}

Based on these results, we combined all mass estimates assuming the
same stellar templates and IMF by means of the Hodges-Lehmann
estimator.  The standard deviation among different methods is a
measure of the systematic uncertainties affecting stellar masses due
to the choice of the assumptions and priors adopted in the fit and
parameter grid sampling.  Such systematics are smaller, by a factor of
2 on average, than the uncertainty due to model degeneracy and scatter
in photometric redshifts.  Among these, the latter dominate over model
degeneracy. However, this can only be tested in sources with robust,
spectroscopic redshifts, and at the same time model degeneracies are
expected to become more important for faint galaxies (lacking
spectroscopic observations) due to the larger photometric errors.

Finally, we compare CANDELS stellar masses with those released by the
3D-HST team. The agreement is satisfying at a first-order glance, but
we observe a negative median offset ($\sim$-0.1dex) in both fields and
a curved trend in UDS. The offset disappears in GOODS-S and is reduced
in UDS when 3D-HST results are compared with the method adopting the
closest assumptions to the 3D-HST team.  While we cannot say whether
or not any one technique is biased, our tests with mock catalogs in
M15 suggests that the median mass is less susceptible to modeling
uncertainties than the results from any one code. The different
behavior of GOODS-S and UDS in the comparison to 3D-HST also
illustrates that differences in photometric quality (number of filters
and S/N) can affect not only the scatter but also the biases between
methods (the poorer quality dataset available in UDS compared to the
more accurate photometry in GOODS-S is more susceptible to biases). In
any case, we show that the larger fraction of objects in the tails of
the $\log (M_*^{\rm MEDIAN}/M_*^{\rm 3D-HST})$ distributions are at
least partially explained by differences in the photometric and
redshift catalogs.

\begin{acknowledgements}
 
We thank the referee for helpful comments.  
PS, AF, MC, RA, KB and EM acknowledge the contribution of the EC FP7 SPACE
  project ASTRODEEP (Ref.No: 312725). PS also acknowledges the grant
  ASI I/005/11/0. SL acknowledges the support by the National Research
  Foundation of Korea (NRF) grant, No. 2008-0060544, funded by the
  Korea government (MSIP). This work is based in part on observations (program
  GO-12060) made with the NASA/ESA {\it Hubble Space Telescope}, which
  is operated by the Association of Universities for Research in
  Astronomy, Inc, under NASA contract NAS5-26555.  This work is also
  based in part on observations made with the {\it Spitzer Space
    Telescope}, which is operated by the Jet Propulsion Laboratory,
  California Institute of Technology under NASA contract 1407.
This work uses data from the following ESO programs: 60.A-9284,
181.A0717, LP 186.A-0898 and 085.A-0961.

\end{acknowledgements}
\bibliographystyle{apj}
%\bibliography{biblio}

\clearpage

 \begin{appendix}

\section{Magellan/IMACS spectroscopy in UDS} \label{app:magellan}

As first discussed in Section~\ref{sec:redshifts}, our analysis and
stellar mass catalog for the UDS field includes redshift measurements
derived from spectroscopic observations in the CANDELS/UDS field using
the Inamori-Magellan Areal Camera and Spectrograph
\citep[IMACS,][]{dressler11} on the Magellan Baade 6.5-meter
telescope. The Magellan/IMACS spectroscopic sample includes a total of
$475$ unique sources, spread over $4$ slitmasks and covering an area
slightly larger than the CANDELS {\it HST}/WFC3-IR footprint in the
UDS. The observations were conducted on the nights of December 30-31,
2010 (UT), with total exposure time of roughly $5400$~seconds per
slitmask ($3~\times~1800$~s with no dithering performed). Immediately
following each set of science exposures (i.e.,~without moving the
telescope or modifying the instrument configuration), a quartz
flat-field frame and comparison arc spectrum (using He, Ar, Ne) were
taken to account for instrument flexure and detector fringing. Each
slitmask contains on the order of 125 slitlets, with a fixed
slitlength and slitwidth of $8^{\prime\prime}$ and $1^{\prime\prime}$,
respectively. We employed the 300~lines/mm grism (${\rm blaze~angle} =
26.7^{\circ}$) with the clear (or "spectroscopic") filter, which
yields a spectral resolution of $R \sim 1200$ at $7500$\AA.

\begin{figure}
  \centering
  \includegraphics[scale=0.5]{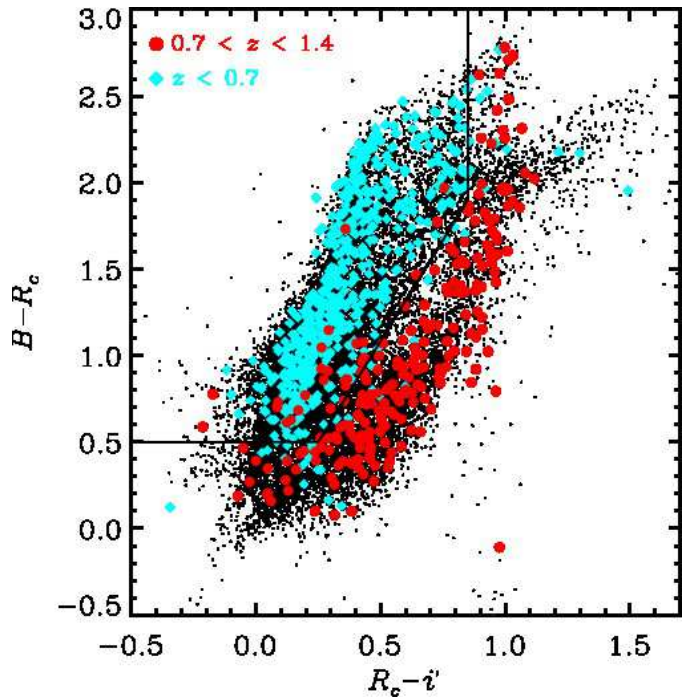}
  \caption{The $B-R_{c}$ versus $R_{c}-I$ color-color distribution for
    all sources with $R_{c} < 23.5$ in the Subaru imaging catalogs of
    \citet{furusawa08}. The cyan and red points correspond to those
    objects with spectroscopic redshifts in the ranges $z < 0.7$ and
    $0.7 < z < 1.4$, respectively. The solid black lines show the
    color cuts employed to prioritize target selection (see
    Equation~\ref{eqn:colorcut}). }
  \label{fig:colorcut}
\end{figure}

The $475$ unique sources in the Magellan/IMACS spectroscopic sample
are drawn from the Subaru optical imaging catalog of
\citet{furusawa08}, which covers the larger $1.22~{\rm degree}^{2}$
Subaru/{\it XMM-Newton} Deep Survey \citep[SXDS,][]{ueda08} field
surrounding the CANDELS/UDS region. We identified spectroscopic
targets according to an $R_{c}$ band limiting magnitude of $R_{c} <
23.5$~(AB), with sources brighter than $R_{c} = 18$ excluded from the
target population.  In an effort to primarily observe galaxies at
intermediate redshift, we prioritized targets according to a
$BR_{c}i^{\prime}$ color selection, closely mirroring that of the
DEEP2 Galaxy Redshift Survey \citep{davis03, davis07, newman13}. The
color-cut, as shown in Figure~\ref{fig:colorcut}, is defined using a
large pool of publicly-available redshifts covering the wider SXDS
field \citep{simpson06, geach07, vB07, smail08} and corresponds to the
following selection criteria:

\begin{align}
(B - R_{c}) &< 0.5~{\rm or} \\
(R_{c} - I) &> 0.85~{\rm or} \\
(B - R_{c}) &< 2.3\bar{3} \times (R_{c} - I)  -0.08\bar{3} .
\label{eqn:colorcut}
\end{align}

\noindent Objects failing these color selection criteria are included
in the sample, but with a lower probability of inclusion in the target
population. In addition, we downweighted those objects with SExtractor
stellarity index of ${\rm CLASS\_STAR} > 0.95$
\citep[i.e.,~stars,][]{bertin96}.

The IMACS spectroscopic observations were reduced using the COSMOS
data reduction pipeline developed at the Carnegie
Observatories\footnote{http://obs.carnegiescience.edu/Code/cosmos}
\citep{dressler11}. For each slitlet, COSMOS yields a flat-fielded and
sky-subtracted, two-dimensional spectrum, with wavelength calibration
performed by fitting to the arc lamp emission lines. One-dimensional
spectra were extracted and redshifts were measured from the reduced
spectra using additional software developed as part of the DEEP2 and
DEEP3 Galaxy Redshift Surveys \citep{newman13, cooper11, cooper12a}
and adapted for use with IMACS as part of the Arizona CDFS Environment
Survey \citep{cooper12b} and as part of the spectroscopic follow-up of
the Red-Sequence Cluster Survey \citep[RCS][zRCS; Yan et al.~in
preparation]{gladders05}. A detailed description of the DEEP2
reduction packages (SPEC2D and SPEC1D) is presented by
\cite{cooper12c} and \cite{newman13}.

\begin{figure}
  \centering
  \includegraphics[scale=0.5]{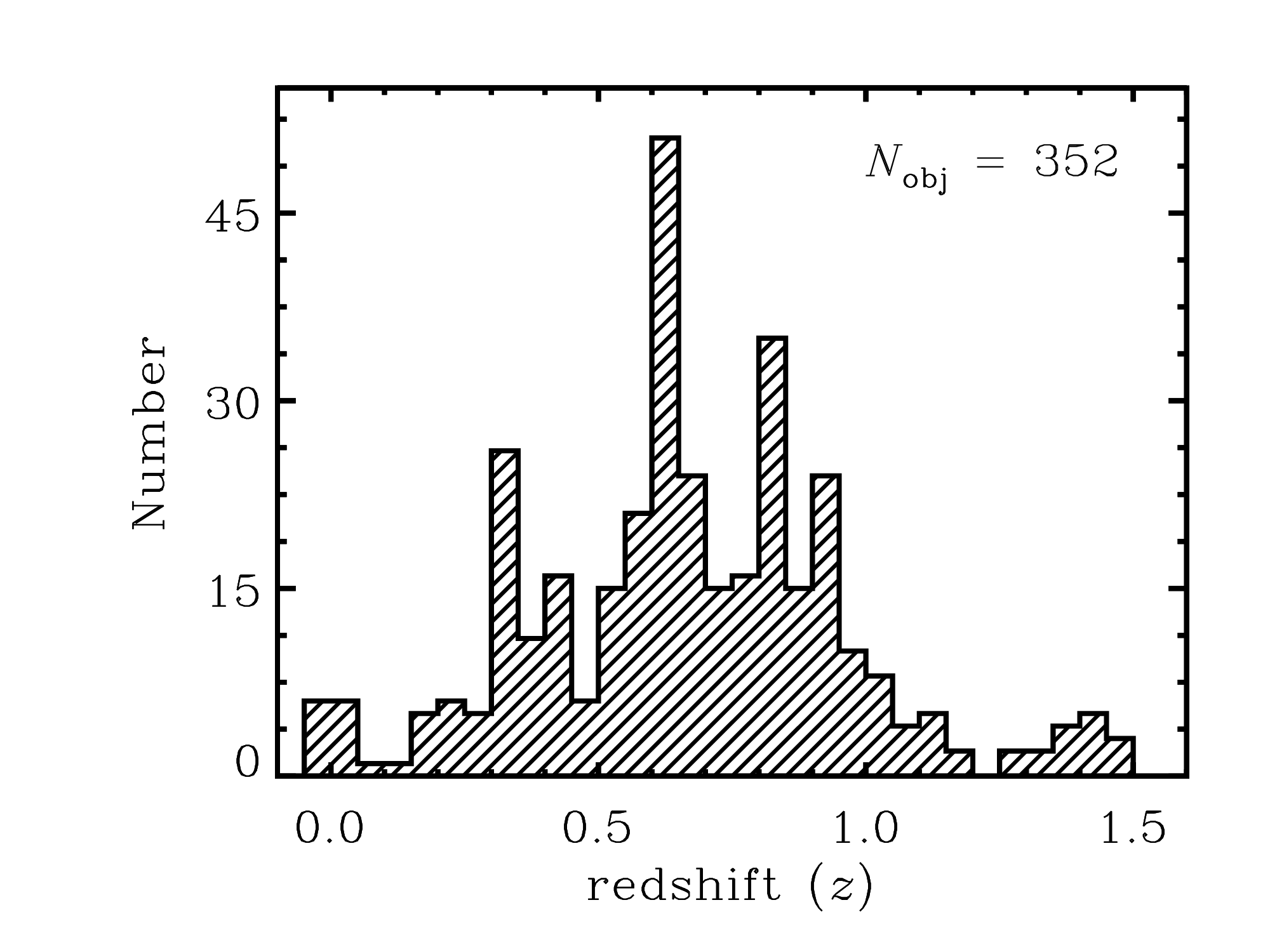}
  \caption{The distribution of the 352 unique, secure ($Q = -1$, $3$,
    $4$) redshifts included in the Magellan/IMACS redshift catalog. }
  \label{fig:dndz}
\end{figure}

All spectra were visually inspected by M.~Cooper, with a quality code
($Q$) assigned corresponding to the accuracy of the redshift value ---
$Q = -1, 3, 4$ denote secure redshifts, with $Q = -1$ corresponding to
stellar sources and $Q = 3, 4$ denoting secure galaxy redshifts (see
Table~\ref{tab:imacs}). Confirmation of multiple spectral features was
generally required to assign a quality code of $Q = 3$ or $4$. Quality
codes of $Q = 1$, $2$ were assigned to observations that yield no
useful redshift information ($Q = 1$) or may possibly yield redshift
information after further analysis or re-reduction of the data ($Q =
2$). For detailed descriptions of the reduction pipeline, redshift
measurement code, and quality assignment process refer to
\citet{wirth04}, \citet{davis07}, and \citet{newman13}.  A redshift
catalog is presented in Table~\ref{tab:imacs}, a subset of which is
listed herein. The entirety of Table~\ref{tab:imacs} appears in the
electronic version of the Journal.  A redshift is only included when
classified as secure ($Q = -1, 3, 4$). The total number of secure
redshifts in the sample is $352$ out of $475$ total, unique
targets. The redshift distribution for this sample, as shown in
Figure~\ref{fig:dndz}, peaks at $z < 1$ with a small tail out to
higher redshift.  Across the $4$ slitmasks, a total of $44$ objects in
the Magellan/IMACS sample were observed more than once. While not a
large sample of repeated observations, these independent spectra
provide a direct means for determining the precision of the redshift
measurements. A comparison of the differences in the redshift
measurements for multiple observations of the same object yields a
redshift precision of $\sigma_{z} \sim 30~{\rm km}~{\rm s}^{-1}$ (for
$Q = -1$, $3$, and $4$).

\begin{deluxetable*}{c c c c c c c c c c c}
\tablewidth{0pt}
\tablecolumns{12}
\tablecaption{\label{tab:imacs} Magellan/IMACS Redshift Catalog}
\tablehead{Object ID\footnotemark[{\it a}] & $\alpha$\footnotemark[{\it b}]
  (J2000) & $\delta$\footnotemark[{\it c}] (J2000) & $R_{c}$\footnotemark[{\it d}] & Mask\footnotemark[{\it e}] & 
  Slit\footnotemark[{\it f}] & MJD\footnotemark[{\it g}] &
  $z$\footnotemark[{\it h}] &
  $z_{\rm helio}$\footnotemark[{\it i}] & $Q$\footnotemark[{\it j}]} 
\startdata
30007451 & 34.212042 & -5.206258 & 21.85 & UDS1 & 1 & 55560.6 &  0.80924 &  0.80915 & 4 \\
30008565 & 34.215329 & -5.214261 & 22.57 & UDS1 & 2 & 55560.6 &  0.81222 &  0.81213 & 4 \\
30009829 & 34.219258 & -5.242188 & 22.84 & UDS1 & 3 & 55560.6 &  0.92968 &  0.92960 & 3 \\
30011686 & 34.226008 & -5.258772 & 22.15 & UDS1 & 4 & 55560.6 &  0.80341 &  0.80332 & 4 \\
\vspace*{-0.1in}
\enddata
\tablecomments{Table \ref{tab:imacs} is presented in its entirety in
  the electronic edition of the Journal. A portion is shown here for
  guidance regarding its form and content.}

\footnotetext[{\it a}]{Object identification number in Subaru imaging
  catalog of \citet{furusawa08}.}

\footnotetext[{\it b}]{Right ascension in decimal degrees from
  \citet{furusawa08}.}

\footnotetext[{\it c}]{Declination in decimal degrees from
  \citet{furusawa08}.}

\footnotetext[{\it d}]{$R$-band magnitude in AB system from
  \citet{furusawa08}.}

\footnotetext[{\it e}]{Name of IMACS slitmask on which object was
  observed.}

\footnotetext[{\it f}]{Number of slit on IMACS slitmask corresponding to
  object.}

\footnotetext[{\it g}]{Modified Julian Date of observation.}

\footnotetext[{\it h}]{Redshift derived from observed spectrum.}

\footnotetext[{\it i}]{Heliocentric-frame redshift.}

\footnotetext[{\it j}]{Redshift quality code (${\rm star} = -1$; 
  ${\rm secure\ redshift} = 3$, $4$; 
  ${\rm unknown} = 1,2$).}

\end{deluxetable*}

\section{Additional stellar mass estimates} \label{app:additional}

In addition to the mass estimates presented in Table \ref{tab:param1},
several teams have provided further results based on different
assumptions, which we present in Table \ref{tab:param3}. We excluded
these estimates from the median computation in order not to overweight
a single method compared to the others. However, we used them to test
how specific parameters affect the best-fit result. Indeed, the
methods listed in Table \ref{tab:param3} offer the advantage of being
based on the very same assumption as their analogues in Table
\ref{tab:param1} except for a single parameter. This makes it possible
to study the effect of such parameter on the output stellar mass by
leaving the other assumptions unchanged.

We list below the differences between the methods presented in
Table~\ref{tab:param3} and those in Table~\ref{tab:param1}.

\begin{itemize} 
\item Method 6a$_\tau^{\rm NEB}$ is completely consistent with Method
  6a$_\tau$ except for inclusion of nebular emission.  Nebular
  emission is treated following \cite{schaerer09}, as already
  presented by \cite{castellano14} and by G15, in a very similar way
  as Method 14a. Briefly, \cite{schaerer09} directly link nebular
  emission to the amount of hydrogen-ionizing photons in the stellar
  spectra \citep{schaerer98} by considering free-free, free-bound, and
  hydrogen two-photon continuum emission. They assume null escape
  fraction, an electron temperature of $10^4$ K, an electron density
  $N_e$=100 cm$^{−3}$, and a 10\% helium numerical abundance relative
  to hydrogen.  Hydrogen lines from the Lyman to the Brackett series
  were included considering Case B recombination, while the relative
  line intensities of He and metals depend on the metallicity
  according to \cite{anders03}.  We refer the reader to the works
  above for more details.

\item Methods 6a$_{\rm del\tau}$ and 6a$_{\rm inv\tau}$ are completely
  consistent with Method 6a$_\tau$ except for the SFH: delayed-$\tau$
  and inverted-$\tau$ models have been used, respectively, instead of
  direct-$\tau$ models.  Delayed-$\tau$ models have a slightly
  different analytic shape ($\psi(t)\propto t^2/\tau \cdot
  exp(-t/\tau)$) compared to Methods 12a, 14a, and 15a presented in
  Table \ref{tab:param1} ($\psi(t)\propto t/\tau^2 \cdot
  exp(-t/\tau)$).

\item Method 10c$^{\rm dust}$ is completely consistent with Method
  10c, but dust reddening has now been included according to a
  Calzetti attenuation law.

\item Method 12a$_\tau$ is consistent with Method 12a except for the
  SFH parameterization (direct-$\tau$ models instead of
  delayed-$\tau$).
\end{itemize}

Moreover, to analyze the effect of the SFH modeling on the mass
estimates, we also take advantage from the results of Method 14a. In
addition to the best-fit results (which we refer to as Method 14a),
this method provides the full set of physical parameters for each of
the four SFH model adopted: constant (Method 14a$_{\rm const}$),
linearly increasing (Method 14a$_{\rm lin}$), delayed-$\tau$ (Method
14a$_{\rm del\tau}$), and direct-$\tau$ models (Method 14a$_\tau$).

\begin{deluxetable*}{cccccc}
\tablecaption{\label{tab:param3} Additional stellar masses in CANDELS
  (not considered for the median computation).}
\tablehead{&\colhead{Method 6a$_\tau^{\rm NEB}$} & \colhead{Method
    6a$_{\rm del\tau}$}
  & \colhead{Method 6a$_{\rm inv\tau}$}
  & \colhead{Method 10c$^{\rm dust}$} & \colhead{Method 12a$_\tau$} }
\tablecolumns{5}
\startdata
{\bf PI} & A. Fontana & A. Fontana& A. Fontana& J. Pforr & T. Wiklind\\
\\[0.1pt]
{\bf fitting method} & min $\chi^2$ & min $\chi^2$ & min $\chi^2$ & min $\chi^2$ & min $\chi^2$ \\
\\[0.1pt]
{\bf code} & zphot\footnote{\cite{giallongo98}, \cite{fontana00}.} & zphot$^a$ &zphot$^a$ & HyperZ\footnote{\cite{bolzonella00}, http://webast.ast.obs-mip.fr/hyperz/.}  & WikZ\footnote{\cite{wiklind08}.} \\
\\[0.1pt]
{\bf stellar templates} & BC03\footnote{\cite{bc03}.} & BC03$^d$& BC03$^d$& BC03$^d$ & M05\footnote{\cite{maraston05}.} \\
\\[0.1pt]
{\bf  IMF} &  Chabrier & Chabrier & Chabrier & Chabrier & Chabrier \\
\\[0.1pt]
{\bf SFH} & $\tau$\footnote{Exponentially decreasing SFH
  (direct-$\tau$ models, see  Section~\ref{sec:massest}).} &
del-$\tau$\footnote{Delayed-$\tau$ models: $\psi(t)\propto t^2/\tau \cdot
  exp(-t/\tau)$.} & inv-$\tau$\footnote{Exponentially increasing SFH
  (inverted-$\tau$ models, see  Section~\ref{sec:massest}).}   &
$\tau$$^f$ + trunc.\footnote{Truncated SFH (see
  Section~\ref{sec:massest}).} + const.\footnote{Constant SFH (see
  Section~\ref{sec:massest}).} & $\tau$$^f$ \\
 \\[0.1pt]
{\bf log ($\tau$/yr)} &8.0--10.2 & 8.0--9.3 &8.0--10.2& 8.5, 9.0, 10.0
& -$\infty$\footnote{The $\tau$ grid starts from 0.0 Gyr in the linear
  space.} -- 9.0 \\
steps\footnote{The number of steps is indicated as the grid size is
  not uniform over the range covered. }   & 9 steps& 20 steps & 9 steps & & 8 steps \\
\\[0.1pt]
{\bf  log ($t_0$/yr)\footnote{$t_0$ is the timescale for truncated SFH
    (see Section~\ref{sec:massest}).}} &  &  & &   8.0, 8.5, 9.0  & \\
\\[0.1pt]
{\bf metallicity [Z$_\odot$] } &  0.02, 0.2, 1, 2.5 &  0.02, 0.2, 1, 2.5  &  0.02, 0.2, 1, 2.5  & 0.2, 0.5, 1, 2.5 & 0.2, 0.4, 1, 2.5 \\
\\[0.1pt]
{\bf log (age/yr)} & 7.0--10.1 & 7.0--10.1 &7.0--10.1 & 8.0--10.3 & 7.7--9.8\\
steps$^l$ & 110 steps & 113 steps & 48 steps & 221 steps & 24 steps\\
\\[0.1pt]
{\bf extinction law} & Calzetti + SMC  & Calzetti + SMC & Calzetti + SMC  & Calzetti & Calzetti \\
\\[0.1pt]
{\bf extinction E(B-V)} & 0.0--1.1& 0.0--1.1& 0.0--1.1& 0.00--0.75 & 0.0--1.0\\
step &0.05& 0.05&0.05& 0.05 & 0.025\\
\\[0.1pt]
{\bf nebular emission }  & yes  & no  & no  & no & no \\
\\[0.1pt]
 {\bf priors } & \footnote{Age must be lower than the age of the
   Universe at the galaxy redshift.} \footnote{Fit
   only fluxes at $\lambda_{RF}<5.5 \mu$m;
   $z_{form}\geq1/\sqrt{\tau}$, where $z_{form}$ is the redshift of
   the onset of the SFH; templates with E(B-V)$>$0.2 and
   age/$\tau$$>$3 or  with  E(B-V)$>$0.1 and Z/Z$_\odot$$<$0.1 or with
   age$>$1Gyr and  Z/Z$_\odot$$<$0.1 are   excluded.} & $^n$ $^o$ &$^n$ $^o$ & $^n$ \footnote{Fit only fluxes at $\lambda_{RF}<2.5 \mu$m.} & $^n$\\
\\[0.1pt]
 {\bf reference } & 4 & 4 & 4 & 5 & 7\\
\enddata
\tablecomments{References. (4) \cite{fontana06}; (5)
  \cite{daddi05,maraston06,pforr12,pforr13}; (7) \cite{wiklind08,wiklind14}.}
\end{deluxetable*}

\section{Notes on the released catalogs} \label{app:masscat}

\begin{figure}[!t]
\centering
  \includegraphics[scale=0.5]{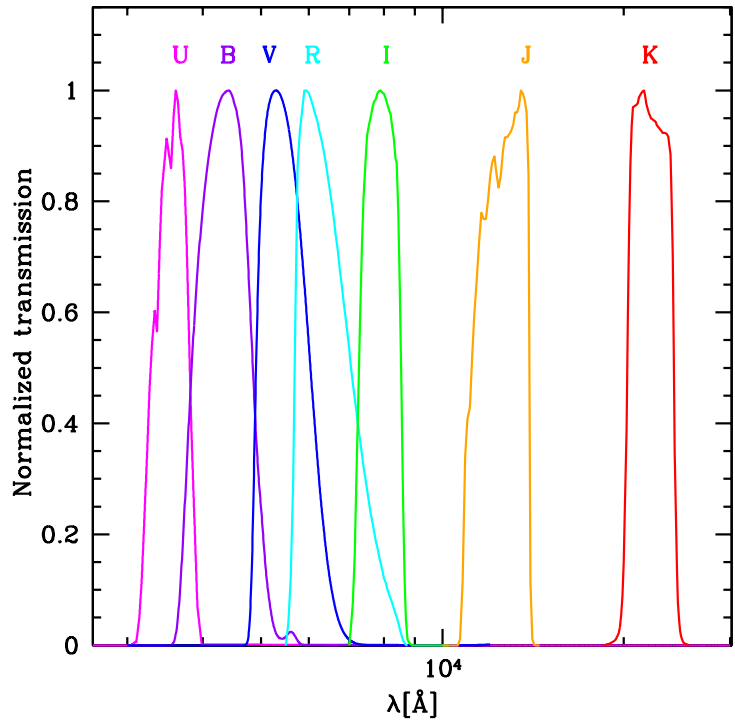}
% \resizebox{0.7\hsize}{!}{
% \includegraphics[angle=0]{filters}}
\caption{Filter curves over which rest-frame magnitudes have been
  computed following Methods 6a$_\tau$, 6a$_\tau^{\rm NEB}$, 6a$_{\rm
    del\tau}$ and 6a$_{\rm inv\tau}$. Filter curves are available as
  ascii files in the electronic version of the Journal.}
\label{fig:filters}
\end{figure}

We report in Tables \ref{tab:masscat} and \ref{tab:physparcat} the
list of columns in the mass catalogs and in the catalogs containing the
other physical parameters, respectively. Fig.\ref{fig:filters} shows
the filter curves over which rest-frame magnitudes have been computed
following Methods 6a$_\tau$, 6a$_\tau^{\rm NEB}$, 6a$_{\rm del\tau}$
and 6a$_{\rm inv\tau}$ (see Table \ref{tab:physparcat}). Filter curves are available as
  ascii files in the electronic version of the Journal.

\begin{deluxetable*}{llll}
\tablecaption{\label{tab:masscat} List of columns in the mass
  catalogs for the GOODS-S (filename GS\_CANDELS\_mass.dat) and
  UDS  (filename UDS\_CANDELS\_mass.dat) fields.}
%\tablehead{\colhead{Column number} & \colhead{Description}& \colhead{Notes}} \\
\tablehead{\colhead{\# column } &\colhead{\# column } & \colhead{Description}& \colhead{Notes} \\
\colhead{GOODS-S} & \colhead{UDS} & \colhead{} & \colhead{} } \\
%\tablehead{\colhead{\# column GOODS-S} & \colhead{Description}& \colhead{Notes}} \\
\startdata
1  & 1 & Designation & \\
2  & 2 & RA &  J2000 \\
3  & 3 & Dec &  J2000 \\
4  & 4 & Observed magnitude in the F160W filter & \\
5  & 5 & Signal-to-noise ratio in the F160W filter & \\
6  & 6 & Photometry flag &  0: ok; $>$0: bad photometry\\
7  & 7 & Spectroscopic star flag &  ?$=$1 spectroscopic star \\
8  & 8 &  Stellarity index from Sextractor on F160W band & \\
9  & 9 & AGN flag &   ?$=$1 Xray AGN \\
10 &  10 & Redshift best estimate &  see catalogs' readme file for details\\
11 &  11 & Spectroscopic redshift & \\
12 &  12 & Quality of spectroscopic redshift &  1: good; 2: intermediate; 3: uncertain \\
13 &  13 & Reference spectroscopic survey & see catalogs' readme file for details\\
14 &  14 & Photometric redshift &  see catalogs' readme file for details \\
15 &  15 & Lower photo-z 68\% confidence limit  & \\
16 &  16 & Upper photo-z 68\% confidence limit  & \\
17 &  17 & Lower photo-z 95\% confidence limit &  \\
18 &  18 & Upper photo-z 95\% confidence limit &  \\
19 &  & Photometric redshift for AGNs &  ?$=$-99  for non AGN sources \\
20 & 19 & CANDELS reference median stellar mass  & $M_*^{\rm MEDIAN}$, see text $[M_\odot]$\\
21 &  20  & Standard deviation on the previous &  $\sigma_{M,CANDELS}$ $[M_\odot]$ \\
22 &  21 & Median stellar mass including nebular component & see text $[M_\odot]$ \\
23 &  22 & Standard deviation on the previous &  $[M_\odot]$\\ 
24 &  23 & Relative uncertainty due to model degeneracy and photo-z scatter & $\delta_{M,z}$,  see text \\
25 &  24 & Stellar mass from Method 2a$_\tau$ & $[\log M/M_\odot]$ \\
26 &  25 & Stellar mass from Method 2d$_\tau$& $[\log M/M_\odot]$\\
 &  26 & Stellar mass from Method 4b& $[\log M/M_\odot]$\\
27 &  27 & Stellar mass from Method 6a$_\tau$& $[\log M/M_\odot]$\\
28 &  28 & Stellar mass from Method 10c  & $[\log M/M_\odot]$\\
29 &  29 &  Stellar mass from Method 11a$_\tau$ &$[\log M/M_\odot]$\\
30 &  30 &  Stellar mass from Method 12a  & $[\log M/M_\odot]$\\
31 &  31 &  Stellar mass from Method 13a$_\tau$ & $[\log M/M_\odot]$\\
32 &  32 &  Stellar mass from Method 14a & $[\log M/M_\odot]$\\
33 &   &  Stellar mass from Method 15a & $[\log M/M_\odot]$\\
34 & 33  &  Stellar mass from Method 6a$_\tau^{\rm NEB}$  & $[\log M/M_\odot]$\\
35 &  34 &  Stellar mass from Method 6a$_{\rm del\tau}$  & $[\log M/M_\odot]$\\
36 &  35 &  Stellar mass from Method 6a$_{\rm inv\tau}$ & $[\log M/M_\odot]$\\
37 &  36 &  Stellar mass from Method 10c$^{\rm dust}$  & $[\log M/M_\odot]$\\
38 &  37 &  Stellar mass from Method 12a$_\tau$ & $[\log M/M_\odot]$\\
39 &  38 &  Stellar mass from Method 14a$_{\rm const}$  & $[\log M/M_\odot]$\\
40 &  39 &  Stellar mass from Method 14a$_{\rm lin}$ & $[\log M/M_\odot]$\\
41 &  40 &  Stellar mass from Method 14a$_{\rm del\tau}$  & $[\log M/M_\odot]$\\
42 & 41 &  Stellar mass from Method 14a$_\tau$  & $[\log M/M_\odot]$\\
\enddata
\end{deluxetable*}

\begin{deluxetable*}{llll}
\tablecaption{\label{tab:physparcat} List of columns in the catalogs including other physical
  parameters for the GOODS-S (filename GS\_CANDELS\_physpar.dat)
  and   UDS  (filename UDS\_CANDELS\_physpar.dat) fields.}
\tablehead{\colhead{\# column } & \colhead{\# column } & \colhead{Description}& \colhead{Notes} \\
\colhead{GOODS-S} &\colhead{UDS} & \colhead{} & \colhead{} } \\
\startdata
1 & 1 & Designation & \\
2 & 2 & Age from Method 2a$_\tau$ & $[\log t/yr]$ \\
3 & 3 &  $\tau$ from Method 2a$_\tau$ & $[Gyr]$\\
4 &  4 & A$_V$ from Method 2a$_\tau$ & [mag]\\
5 &  5 & SFR from Method 2a$_\tau$& $[M_\odot/yr]$\\
6 &  6 & Reduced  $\chi^2$ from Method 2a$_\tau$ & \\
7 &  7 & Age from Method 2d$_\tau$ & $[\log t/yr]$ \\
8 &  8 & $\tau$ from Method 2d$_\tau$ & $[Gyr]$\\
9 &  9 & A$_V$ from Method 2d$_\tau$ & [mag]\\
10 & 10  & Gas metallicity from Method 2d$_\tau$ & $[Z_\odot]$\\
&  11 & Age from Method 4b & $[\log t/yr]$ \\
&  12 & E(B-V) from Method 4b & [mag]\\
11 & 13   & Lower stellar mass 68\% confidence limit from Method 6a$_\tau$ &$[\log M/M_\odot]$\\
12 &  14 & Upper stellar mass 68\% confidence limit from Method 6a$_\tau$ &$[\log M/M_\odot]$\\
13 & 15   & Age from Method 6a$_\tau$ & $[\log t/yr]$\\
14 &  16  &$\tau$ from Method 6a$_\tau$ & $[Gyr]$\\
15 &  17  & E(B-V) from Method 6a$_\tau$ & [mag]\\
16 &  18  & SFR from Method 6a$_\tau$ & $[M_\odot/yr]$\\
17 &  19  & Gas metallicity from Method 6a$_\tau$ & $[Z_\odot]$\\
18 &  20 & Extinction law from Method 6a$_\tau$ & 1: Calzetti et al. 2000; 2: SMC\\
19 &  21  & Reduced $\chi^2$ from Method 6a$_\tau$ & \\
20 &  22  & Rest-frame luminosity at 1400\AA from Method 6a$_\tau$  & $L_\nu(1400\AA)~ [erg/s/Hz]$\\
21 &  23  & Rest-frame luminosity at 2700\AA from Method 6a$_\tau$  & $L_\nu(2700\AA)~ [erg/s/Hz]$\\
22 &  24  & Rest-frame magnitude in the U band from Method 6a$_\tau$  &  [mag$_{\rm AB}$], see Fig.\ref{fig:filters}\\
23 &  25  & Rest-frame magnitude in the B band  from Method 6a$_\tau$ &  [mag$_{\rm AB}$], see Fig.\ref{fig:filters}\\
24 &  26  & Rest-frame magnitude in the V band from Method 6a$_\tau$  &  [mag$_{\rm AB}$], see Fig.\ref{fig:filters}\\
25 &  27 & Rest-frame magnitude in the R band from Method 6a$_\tau$  & [mag$_{\rm AB}$], see Fig.\ref{fig:filters} \\
26 &  28  & Rest-frame magnitude in the I band  from Method 6a$_\tau$ &  [mag$_{\rm AB}$], see Fig.\ref{fig:filters}\\
27 &  29  & Rest-frame magnitude in the J band  from Method 6a$_\tau$ &  [mag$_{\rm AB}$], see Fig.\ref{fig:filters}\\
28 &  30  & Rest-frame magnitude in the K band from Method 6a$_\tau$&  [mag$_{\rm AB}$], see Fig.\ref{fig:filters}\\
29 &  31 & Age from Method 10c & $[\log t/yr]$\\
30 &  32  & Star formation history from Method 10c & see catalogs' readme file for details\\
31 &  33  &$\tau$ from Method 10c & $[Gyr]$, see catalogs' readme file for details \\
32 &  34 & Gas metallicity from Method 10c & $[Z_\odot]$\\
33 &  35  & Lower stellar mass 99\% confidence limit from Method 11a$_\tau$ &$[\log M/M_\odot]$\\
34 &  36  & Upper stellar mass 99\% confidence limit from Method 11a$_\tau$ &$[\log M/M_\odot]$\\
35 &  37  & Age from Method 11a$_\tau$ & $[\log t/yr]$\\
36 &  38  & SFR from Method 11a$_\tau$ & $[M_\odot/yr]$\\
37 &  39  & Lower stellar mass 68\% confidence limit from Method 12a &$[\log M/M_\odot]$\\
38 &  40  & Upper stellar mass 68\% confidence limit from Method 12a &$[\log M/M_\odot]$\\
39 &  41  & Lower stellar mass 95\% confidence limit from Method 12a &$[\log M/M_\odot]$\\
40 &  42  & Upper stellar mass 95\% confidence limit from Method 12a &$[\log M/M_\odot]$\\
41 &  43  & Age from Method 12a & $[\log t/yr]$\\
42 &  44  &$\tau$ from Method 12a & $[Gyr]$\\
43 &  45  & E(B-V) from Method 12a& [mag]\\
44 &  46  & Gas metallicity from Method 12a & $[Z_\odot]$\\
45 &  47  & Stellar bolometric luminosity corrected for dust & $[\log L/L_\odot]$\\
  &     & extinction from Method 12a & \\
46 &  48  & Reduced $\chi^2$ from Method 12a & see catalogs' readme file for details \\
47 &  49  & Age from Method 13a$_\tau$ & $[\log t/yr]$ \\
48 &   50 & $\tau$ from Method 13a$_\tau$ & $[Gyr]$\\
49 & 51  & A$_V$ from Method 13a$_\tau$ & [mag]\\
50 &  52  & SFR from Method 13a$_\tau$& $[M_\odot/yr]$\\
51 &  53  & Reduced  $\chi^2$ from Method 13a$_\tau$ & \\
52 &  54  & Age from Method 14a & $[\log t/yr]$\\
53 &  55  & Star formation history from Method 14a & see catalogs' readme file for details\\
54 &  56 & $\tau$ from Method 14a & $[Gyr]$, see catalogs' readme file for details \\
55 &  57  & E(B-V) from Method 14a & [mag]\\
56 &  58  & SFR from Method 14a& $[M_\odot/yr]$\\
57 &   59 & Fit quality from Method 14a& 1:best; 2:good; others:bad\\
58 &    & Age from Method 15a & $[\log t/yr]$ \\
59 &    & $\tau$ from Method 15a & $[Gyr]$\\
60 &    & E(B-V) from Method 15a & [mag]\\
61 &   & Gas metallicity from Method 15a & $[Z_\odot]$\\
62 &  60  & Age from Method 6a$_\tau^{\rm NEB}$ & $[\log t/yr]$\\
63 &  61  & $\tau$ from Method 6a$_\tau^{\rm NEB}$ & $[Gyr]$\\
64 &  62  & E(B-V) from Method 6a$_\tau^{\rm NEB}$ & [mag]\\
65 &  63  & SFR from Method 6a$_\tau^{\rm NEB}$ & $[M_\odot/yr]$\\
66 &  64  & Gas metallicity from Method 6a$_\tau^{\rm NEB}$ & $[Z_\odot]$\\
67 &  65 & Extinction law from Method 6a$_\tau^{\rm NEB}$ & 1: Calzetti et al. 2000; 2: SMC\\
68 &  66  & Reduced $\chi^2$ from Method 6a$_\tau^{\rm NEB}$ & \\
\enddata
\end{deluxetable*}

\begin{deluxetable*}{llll}
\tablenum{5}
\tablecaption{%\label{tab:physparcat} 
  ({\it continued})}
\tablehead{\colhead{\# column } & \colhead{\# column } & \colhead{Description}& \colhead{Notes} \\
\colhead{GOODS-S} &\colhead{UDS} & \colhead{} & \colhead{} } \\
\startdata
69 &  67  & Rest-frame luminosity at 1400\AA from Method 6a$_\tau^{\rm NEB}$  & $L_\nu(1400\AA)~ [erg/s/Hz]$\\
70 &  68  & Rest-frame luminosity at 2700\AA from Method 6a$_\tau^{\rm NEB}$  & $L_\nu(2700\AA)~ [erg/s/Hz]$\\
71 &  69  & Rest-frame magnitude in the U band from Method 6a$_\tau^{\rm NEB}$  & [mag$_{\rm AB}$], see Fig.\ref{fig:filters}\\
72 &  70  & Rest-frame magnitude in the B band  from Method6a$_\tau^{\rm NEB}$ & [mag$_{\rm AB}$], see Fig.\ref{fig:filters} \\
73 &  71  & Rest-frame magnitude in the V band from Method 6a$_\tau^{\rm NEB}$  & [mag$_{\rm AB}$],  see Fig.\ref{fig:filters}\\
74 &  72  & Rest-frame magnitude in the R band from Method 6a$_\tau^{\rm NEB}$  & [mag$_{\rm AB}$],  see Fig.\ref{fig:filters}\\
75 &  73  & Rest-frame magnitude in the I band  from Method 6a$_\tau^{\rm NEB}$ & [mag$_{\rm AB}$],  see Fig.\ref{fig:filters}\\
76 &  74  & Rest-frame magnitude in the J band  from Method 6a$_\tau^{\rm NEB}$ & [mag$_{\rm AB}$], see Fig.\ref{fig:filters}\\
77 &  75  & Rest-frame magnitude in the K band from Method 6a$_\tau^{\rm NEB}$  & [mag$_{\rm AB}$], see Fig.\ref{fig:filters}\\
78 &76 &  Age from Method 6a$_{\rm del\tau}$ & $[\log t/yr]$\\
79 &77 &  $\tau$ from Method 6a$_{\rm del\tau}$ & $[Gyr]$\\
80 &78 &  E(B-V) from Method 6a$_{\rm del\tau}$ & [mag]\\
81 &79 &  SFR from Method 6a$_{\rm del\tau}$ & $[M_\odot/yr]$\\
82 &80 &  Gas metallicity from Method 6a$_{\rm del\tau}$ & $[Z_\odot]$\\
83 &81 &  Extinction law from Method 6a$_{\rm del\tau}$ & 1: Calzetti et al. 2000; 2: SMC\\
84 & 82&  Reduced $\chi^2$ from Method 6a$_{\rm del\tau}$ & \\
85 & 83&  Rest-frame luminosity at 1400\AA from Method 6a$_{\rm del\tau}$  & $L_\nu(1400\AA)~ [erg/s/Hz]$\\
86 & 84&  Rest-frame luminosity at 2700\AA from Method 6a$_{\rm del\tau}$  & $L_\nu(2700\AA)~ [erg/s/Hz]$\\
87 & 85&  Rest-frame magnitude in the U band from Method 6a$_{\rm del\tau}$  &  [mag$_{\rm AB}$], see Fig.\ref{fig:filters}\\
88 & 86&  Rest-frame magnitude in the B band  from Method 6a$_{\rm del\tau}$ &  [mag$_{\rm AB}$], see Fig.\ref{fig:filters}\\
89 & 87&  Rest-frame magnitude in the V band from Method 6a$_{\rm del\tau}$  &  [mag$_{\rm AB}$], see Fig.\ref{fig:filters}\\
90 & 88&  Rest-frame magnitude in the R band from Method 6a$_{\rm  del\tau}$  &  [mag$_{\rm AB}$], see Fig.\ref{fig:filters}\\
91 & 89&  Rest-frame magnitude in the I band  from Method 6a$_{\rm  del\tau}$ &  [mag$_{\rm AB}$], see Fig.\ref{fig:filters} \\
92 & 90&  Rest-frame magnitude in the J band  from Method 6a$_{\rm  del\tau}$ &  [mag$_{\rm AB}$], see Fig.\ref{fig:filters} \\
93 & 91&  Rest-frame magnitude in the K band from Method 6a$_{\rm del\tau}$  &  [mag$_{\rm AB}$], see Fig.\ref{fig:filters}\\
94 & 92&  Age from Method 6a$_{\rm inv\tau}$ & $[\log t/yr]$\\
95 & 93&  $\tau$ from Method 6a$_{\rm inv\tau}$ & $[Gyr]$\\
96 & 94&  E(B-V) from Method 6a$_{\rm inv\tau}$ & [mag]\\
97 & 95&  SFR from Method 6a$_{\rm inv\tau}$ & $[M_\odot/yr]$\\
98 & 96&  Gas metallicity from Method 6a$_{\rm inv\tau}$ & $[Z_\odot]$\\
99 & 97&  Extinction law from Method 6a$_{\rm inv\tau}$ & 1: Calzetti et al. 2000; 2: SMC\\
100 & 98&  Reduced $\chi^2$ from Method 6a$_{\rm inv\tau}$ & \\
101 & 99&  Rest-frame luminosity at 1400\AA from Method 6a$_{\rm inv\tau}$  & $L_\nu(1400\AA)~ [erg/s/Hz]$\\
102 & 100&  Rest-frame luminosity at 2700\AA from Method 6a$_{\rm inv\tau}$  & $L_\nu(2700\AA)~ [erg/s/Hz]$\\
103 & 101&  Rest-frame magnitude in the U band from Method 6a$_{\rm inv\tau}$  & see Fig.\ref{fig:filters} [mag$_{\rm AB}$]\\
104 & 102&  Rest-frame magnitude in the B band  from Method 6a$_{\rm inv\tau}$ & see Fig.\ref{fig:filters} [mag$_{\rm AB}$]\\
105 & 103&  Rest-frame magnitude in the V band from Method 6a$_{\rm inv\tau}$  & see Fig.\ref{fig:filters} [mag$_{\rm AB}$]\\
106 & 104&  Rest-frame magnitude in the R band from Method 6a$_{\rm inv\tau}$  & see Fig.\ref{fig:filters} [mag$_{\rm AB}$]\\
107 & 105&  Rest-frame magnitude in the I band  from Method 6a$_{\rm inv\tau}$ & see Fig.\ref{fig:filters} [mag$_{\rm AB}$]\\
108 & 106&  Rest-frame magnitude in the J band  from Method 6a$_{\rm inv\tau}$ & see Fig.\ref{fig:filters} [mag$_{\rm AB}$]\\
109 & 107&  Rest-frame magnitude in the K band from Method 6a$_{\rm inv\tau}$  & see Fig.\ref{fig:filters} [mag$_{\rm AB}$]\\
110 & 108&  Age from Method 10c$^{\rm dust}$ & $[\log t/yr]$\\
111 & 109&  Star formation history from Method 10c$^{\rm dust}$ & see catalogs' readme file for details\\
112 & 110&  $\tau$ from Method 10c$^{\rm dust}$ & $[Gyr]$, see catalogs' readme file for details \\
113 & 111& Gas metallicity from Method 10c$^{\rm dust}$ & $[Z_\odot]$\\
114 & 112&  Age from Method 12a$_\tau$ & $[\log t/yr]$\\
115 & 113&  $\tau$ from Method 12a$_\tau$ & $[Gyr]$\\
116 & 114&  E(B-V) from Method 12a$_\tau$ & [mag]\\
117 & 115&  Gas metallicity from Method 12a$_\tau$ & $[Z_\odot]$\\
118 &  116  & Stellar bolometric luminosity corrected for dust & $[\log L/L_\odot]$\\
  &     & extinction from Method 12a$_\tau$ & \\
119 & 117&  Reduced $\chi^2$ from Method 12a$_\tau$ & see catalogs' readme file for details \\
120 & 118&  Age from Method 14a$_{\rm const}$ & $[\log t/yr]$\\
121 & 119&  E(B-V) from Method 14a$_{\rm const}$ & [mag]\\
122 & 120&  SFR from Method 14a$_{\rm const}$ & $[M_\odot/yr]$\\
123 & 121&  Fit quality from Method 14a$_{\rm const}$ & 1:best; 2:good; others:bad\\
124 & 122&  Age from Method 14a$_{\rm lin}$ & $[\log t/yr]$\\
125 & 123&  E(B-V) from Method 14a$_{\rm lin}$ & [mag]\\
126 & 124&  SFR from Method 14a$_{\rm lin}$ & $[M_\odot/yr]$\\
127 & 125&  Fit quality from Method 14a$_{\rm lin}$ & 1:best; 2:good; others:bad\\
128 & 126&  Age from Method 14a$_{\rm del\tau}$ & $[\log t/yr]$\\
129 & 127&  $\tau$ from Method 14a & $[Gyr]$, see catalogs' readme file for details \\
130 & 128&  E(B-V) from Method 14a$_{\rm del\tau}$ & [mag]\\
131 & 129&  SFR from Method 14a$_{\rm del\tau}$ & $[M_\odot/yr]$\\
132 & 130&  Fit quality from Method 14a$_{\rm del\tau}$ & 1:best; 2:good; others:bad\\
133 & 131&  Age from Method 14a$_\tau$ & $[\log t/yr]$\\
134 & 132&  $\tau$ from Method 14a & $[Gyr]$, see catalogs' readme file for details \\
135 & 133&  E(B-V) from Method 14a$_\tau$ & [mag]\\
136 & 133&  SFR from Method 14a$_\tau$ & $[M_\odot/yr]$\\
137 & 135&  Fit quality from Method 14a$_\tau$ & 1:best; 2:good; others:bad\\
\enddata
\end{deluxetable*}

\end{appendix}

% \tablecaption{\label{tab:param3} Additional stellar masses in CANDELS
%   (not considered for the median computation).}
% \tablehead{&\colhead{Method 6a$_\tau^{\rm NEB}$} & \colhead{Method
%     6a$_{\rm del\tau}$}
%   & \colhead{Method 6a$_{\rm inv\tau}$}
%   & \colhead{Method 10c$^{\rm dust}$} & \colhead{Method 12a$_\tau$} }
% \tablecolumns{5}

\end{document}